\newcommand\apjcls{1}
\newcommand\aastexcls{2}
\newcommand\othercls{3}

\newcommand\papercls{\aastexcls}
\documentclass[tighten, times, twocolumn]{aastex62}  % ApJ look-alike
\if\papercls \apjcls
\usepackage{apjfonts}
\else\if\papercls \othercls
\usepackage{epsfig}
\usepackage{margin}
\usepackage{times}
\fi\fi
\usepackage{ifthen}
\usepackage{natbib}
\usepackage{amssymb, amsmath}
\usepackage{appendix}
\usepackage{etoolbox}
\usepackage[T1]{fontenc}
\usepackage{paralist}

\if\papercls \apjcls
\newcommand\aas{\ref@jnl{AAS Meeting Abstracts}}% *** added by jh
\newcommand\dps{\ref@jnl{AAS/DPS Meeting Abstracts}}% *** added by jh
\newcommand\maps{\ref@jnl{MAPS}}% *** added by jh
\else\if\papercls \othercls
\usepackage{astjnlabbrev-jh}
\fi\fi

\if\papercls \aastexcls
\hypersetup{citecolor=blue, % color for \cite{...} links
            linkcolor=blue, % color for \ref{...} links
            menucolor=blue, % color for Acrobat menu buttons
            urlcolor=blue}  % color for \url{...} links
\else
\usepackage[%pdftex,      % hyper-references for pdflatex
bookmarks=true,           %%% generate bookmarks ...
bookmarksnumbered=true,   %%% ... with numbers
colorlinks=true,          % links are colored
citecolor=blue,           % color for \cite{...} links
linkcolor=blue,           % color for \ref{...} links
menucolor=blue,           % color for Acrobat menu buttons
urlcolor=blue,            % color for \url{...} links
linkbordercolor={0 0 1},  %%% blue frames around links
pdfborder={0 0 1},
frenchlinks=true]{hyperref}
\fi

\if\papercls \othercls

\else

\fi

\providecommand{\adsurl}[1]{\href{#1}{ADS}}

\makeatletter
\patchcmd{\NAT@citex}
  {\@citea\NAT@hyper@{%
     \NAT@nmfmt{\NAT@nm}%
     \hyper@natlinkbreak{\NAT@aysep\NAT@spacechar}{\@citeb\@extra@b@citeb}%
     \NAT@date}}
  {\@citea\NAT@nmfmt{\NAT@nm}%
   \NAT@aysep\NAT@spacechar\NAT@hyper@{\NAT@date}}{}{}

\patchcmd{\NAT@citex}
  {\@citea\NAT@hyper@{%
     \NAT@nmfmt{\NAT@nm}%
     \hyper@natlinkbreak{\NAT@spacechar\NAT@@open\if*#1*\else#1\NAT@spacechar\fi}%
       {\@citeb\@extra@b@citeb}%
     \NAT@date}}
  {\@citea\NAT@nmfmt{\NAT@nm}%
   \NAT@spacechar\NAT@@open\if*#1*\else#1\NAT@spacechar\fi\NAT@hyper@{\NAT@date}}
  {}{}
\makeatother

\makeatletter
\DeclareRobustCommand{\lowcase}[1]{\@lowcase#1\@nil}
\def\@lowcase#1\@nil{\if\relax#1\relax\else\MakeLowercase{#1}\fi}
\makeatother

\DeclareSymbolFont{UPM}{U}{eur}{m}{n}
\DeclareMathSymbol{\umu}{0}{UPM}{"16}
\let\oldumu=\umu
\renewcommand\umu{\ifmmode\oldumu\else\math{\oldumu}\fi}

\if\papercls \othercls

\else

\fi

\let\oldsim=\sim
\renewcommand\sim{\ifmmode\oldsim\else\math{\oldsim}\fi}
\let\oldpm=\pm
\renewcommand\pm{\ifmmode\oldpm\else\math{\oldpm}\fi}
\newcommand\by{\ifmmode\times\else\math{\times}\fi}
\newbox{\wdbox}
\renewcommand\c{\setbox\wdbox=\hbox{,}\hspace{\wd\wdbox}}
\renewcommand\i{\setbox\wdbox=\hbox{i}\hspace{\wd\wdbox}}

\newcount\timect
\newcount\hourct
\newcount\minct
\newcommand\now{\timect=\time \divide\timect by 60
         \hourct=\timect \multiply\hourct by 60
         \minct=\time \advance\minct by -\hourct
         \number\timect:\ifnum \minct < 10 0\fi\number\minct}

\catcode`@=11

\newcommand\comment[1]{}

\newcommand\commenton{\catcode`\%=14}

\renewcommand\math[1]{$#1$}
\newcommand\mathshifton{\catcode`\$=3}

\let\atab=&
\newcommand\atabon{\catcode`\&=4}

\let\oldmsp=\sp
\let\oldmsb=\sb
\def\sp#1{\ifmmode
           \oldmsp{#1}%
         \else\strut\raise.85ex\hbox{\scriptsize #1}\fi}
\def\sb#1{\ifmmode
           \oldmsb{#1}%
         \else\strut\raise-.54ex\hbox{\scriptsize #1}\fi}
\newbox\@sp
\newbox\@sb
\def\sbp#1#2{\ifmmode%
           \oldmsb{#1}\oldmsp{#2}%
         \else
           \setbox\@sb=\hbox{\sb{#1}}%
           \setbox\@sp=\hbox{\sp{#2}}%
           \rlap{\copy\@sb}\copy\@sp
           \ifdim \wd\@sb >\wd\@sp
             \hskip -\wd\@sp \hskip \wd\@sb
           \fi
        \fi}
\def\msp#1{\ifmmode
           \oldmsp{#1}
         \else \math{\oldmsp{#1}}\fi}
\def\msb#1{\ifmmode
           \oldmsb{#1}
         \else \math{\oldmsb{#1}}\fi}

\def\supon{\catcode`\^=7}

\def\subon{\catcode`\_=8}

\def\supsubon{\supon \subon}

\newcommand\actcharon{\catcode`\~=13}

\newcommand\paramon{\catcode`\#=6}

\comment{And now to turn us totally on and off...}

\newcommand\reservedcharson{ \commenton  \mathshifton  \atabon  \supsubon 
                             \actcharon  \paramon}

\catcode`@=12
\reservedcharson

\if\papercls \apjcls

\else

\fi

\newcommand\chisq{\ifmmode{\chi\sp{2}}\else\math{\chi\sp{2}}\fi}
\newcommand\redchisq{\ifmmode{ \chi\sp{2}\sb{\rm red}}
                    \else\math{\chi\sp{2}\sb{\rm red}}\fi}
\newcommand\Teq{\ifmmode{T\sb{\rm eq}}\else$T$\sb{eq}\fi}
\newcommand\mjup{\ifmmode{M\sb{\rm Jup}}\else$M$\sb{Jup}\fi}
\newcommand\rjup{\ifmmode{R\sb{\rm Jup}}\else$R$\sb{Jup}\fi}
\newcommand\msun{\ifmmode{M\sb{\odot}}\else$M\sb{\odot}$\fi}
\newcommand\rsun{\ifmmode{R\sb{\odot}}\else$R\sb{\odot}$\fi}
\newcommand\mearth{\ifmmode{M\sb{\oplus}}\else$M\sb{\oplus}$\fi}
\newcommand\rearth{\ifmmode{R\sb{\oplus}}\else$R\sb{\oplus}$\fi}
\shorttitle{Narrow Fe-K$\alpha$ Reverberation Mapping of Changing-Look AGN}
\shortauthors{Noda {\em et al.}}

\begin{document}

%\title{Narrow Fe-K$\alpha$ Reverberation Mapping Unveils the Broad-Line Region Structure in a Changing-Look Active Galactic Nucleus}
\title{Narrow Fe-K$\alpha$ Reverberation Mapping Unveils the Deactivated Broad-Line Region in a Changing-Look Active Galactic Nucleus}

%% AUTHOR/INSTITUTIONS FOR AASTEX6.1:
\author{Hirofumi Noda}
\affiliation{Department of Earth and Space Science, Graduate School of Science, Osaka University, 1-1 Machikaneyama, Toyonaka, Osaka 560-0043, Japan}

\author{Taisei Mineta}
\affiliation{Department of Earth and Space Science, Graduate School of Science, Osaka University, 1-1 Machikaneyama, Toyonaka, Osaka 560-0043, Japan}

\author{Takeo Minezaki}
\affiliation{Institute of Astronomy, School of Science, the University of Tokyo, 2-21-1 Osawa, Mitaka, Tokyo 181-0015, Japan}

\author{Hiroaki Sameshima}
\affiliation{Institute of Astronomy, School of Science, the University of Tokyo, 2-21-1 Osawa, Mitaka, Tokyo 181-0015, Japan}

\author{Mitsuru Kokubo}
\affiliation{Department of Astrophysical Sciences, Princeton University, Princeton, New Jersey 08544,USA}

\author{Taiki Kawamuro}
\affiliation{RIKEN Cluster for Pioneering Research, 2-1 Hirosawa, Wako, Saitama 351-0198, Japan}

\author{Satoshi Yamada}
\affiliation{RIKEN Cluster for Pioneering Research, 2-1 Hirosawa, Wako, Saitama 351-0198, Japan}

\author{Takashi Horiuchi }
\affiliation{Institute of Astronomy, School of Science, the University of Tokyo, 2-21-1 Osawa, Mitaka, Tokyo 181-0015, Japan}

\author{Hironori Matsumoto}
\affiliation{Department of Earth and Space Science, Graduate School of Science, Osaka University, 1-1 Machikaneyama, Toyonaka, Osaka 560-0043, Japan}

\author{Makoto Watanabe}
\affiliation{Department of Physics, Okayama University of Science, 1-1 Ridai-cho, Kita-ku, Okayama, Okayama 700-0005, Japan}

\author{Kumiko Morihana}
\affiliation{Subaru Telescope, National Astronomical Observatory of Japan, 650 North A'ohoku Place, Hilo, HI 96720, USA}

\author{Yoichi Itoh}
\affiliation{Nishi-harima Astronomical Observatory, Center for Astronomy, University of Hyogo, 407-2 Nichigaichi, Sayo-cho, Sayo, Hyogo 679-5313, Japan}

\author{Koji S. Kawabata}
\affiliation{Hiroshima Astrophysical Science Center, Hiroshima University, Higashi-Hiroshima, Hiroshima 739-8526, Japan}

\author{Yasushi Fukazawa}
\affiliation{Department of Physical Science, Hiroshima University, 1-3-1 Kagamiyama, Higashi-Hiroshima, Hiroshima 739-8526, Japan}

%% AUTHOR/INSTITUTIONS FOR EMULATE APJ:
% \author{Patricio~E.~Cubillos\altaffilmark{1,2},
% Joseph~Harrington\altaffilmark{1},
% and
% Third~Author\altaffilmark{1}
% }
% \affil{\sp{1} Planetary Sciences Group, Department of
%               Physics, University of Central Florida, Orlando, FL 32816-2385\\
%        \sp{2} Space Research Institute, Austrian Academy of Sciences,
%               Schmiedlstrasse 6, A-8042, Graz, Austria}

\email{noda@ess.sci.osaka-u.ac.jp}

% %% Extra info for aastex:
% \received{Yesterday}
% \revised{Today}
% \accepted{Tonight}
% \published{Tomorrow}
% \submitjournal{AASJournal}

\begin{abstract}
``Changing-look active galactic nuclei'' (CLAGNs) are known to change their apparent types between types 1 and 2, usually accompanied by a drastic change in their luminosity on timescales of years. 
However, it is still unclear whether materials in broad-line regions (BLRs) in CLAGNs appear and disappear during the type-transition or remain at the same location while the line production is simply activated or deactivated.   
Here we present our X-ray--optical monitoring results of a CLAGN, NGC~3516, by \textit{Suzaku}, \textit{Swift}, and ground telescopes, with our primary focus on the narrow Fe-K$\alpha$ emission line, which is an effective probe of the BLR materials. 
We detected significant variations of the narrow Fe-K$\alpha$ line on a timescale of tens~of~days during the type-2~(faint) phase in 2013--2014, and conducted ``narrow Fe-K$\alpha$ reverberation mapping,'' comparing its flux variation with those of the X-ray continuum from a corona and $B$-band continuum from an accretion disk. 
We derived, as a result, a time lag of $10.1^{+5.8}_{-5.6}$~days~($1\sigma$ errors) for the Fe-K$\alpha$ line behind the continuum, which is consistent with the location of the BLR determined in optical spectroscopic reverberation mapping during the type-1~(bright) phase. 
This finding shows that the BLR materials remained at the same location without emitting optical broad-lines during the type-2 phase. 
Considering the drastic decrease of the radiation during the type-transition, our result is possibly inconsistent with the hotly-discussed formation models of the BLR which propose that the radiative pressure from an accretion disk should be the main driving force.  
\end{abstract}
% http://journals.aas.org/authors/keywords2013.html

%--------------------Introduction---------------------
\section{Introduction}

%---------------------------------------------------------

Active Galactic Nuclei (AGNs) are characterized in the optical bands with so-called broad and narrow emission lines with velocity widths of $v_{\rm w}\gtrsim 1000$~km~s$^{-1}$ and several hundreds km~s$^{-1}$, respectively. Whereas the latter are observed in most AGNs, some AGNs almost completely lack the former. Accordingly AGNs are commonly categorized into two types of 1 and 2, depending on whether they show or not, respectively, broad lines (e.g., \citealt{1943ApJ....97...28S}; \citealt{1974ApJ...192..581K}). 
Type-1 AGNs are further categorized into type 1.2, 1.5, 1.8, and 1.9  according to the relative strength of the broad H$\beta$ to broad H$\alpha$ lines (e.g., \citealt{1977ApJ...215..733O}; \citealt{1981ApJ...249..462O}). 
In  types 1.8--1.9, many of the strong optical broad-lines that are commonly observed in proper type-1 AGNs are absent,  and hence their  characters are likely to be closer to those in type 2  than type 1. 
 The discovery of hidden broad emission-lines  in optical polarized spectra from type-2 AGNs (e.g., \citealt{1985ApJ...297..621A}; \citealt{1990ApJ...355..456M}) led to a hypothesis that all AGNs  have identical structures consisting of, radially from the center to  outer regions, a supermassive black hole (SMBH), X-ray corona, accretion disk, broad-line region (BLR), dusty torus, and narrow-line region (NLR). The hypothesis  is now widely accepted and is known as the AGN unified model (\citealt{1993ARA&A..31..473A}; \citealt{1995PASP..107..803U}). 
In the unified model, the two types  are explained  in terms of the difference  in the viewing angles, where zero degrees is defined  in the direction of the polar axis; 
type-1 AGNs  have low viewing angles  such that their BLRs  are visible without obscuration by dusty tori,  whereas type-2 AGNs  have large viewing angles  such that their BLRs are obscured by dusty tori. 

Although the unified model has been successful  in explaining the characteristics of many AGNs for  the past three decades,  an increasing number of observational results  that apparently challenge the unified model  have been emerging. 
One of the most perplexing and thus hotly debated  facts is that some AGNs  change their types between type 1 and 1.9--2, usually accompanied by  luminosity variation by orders of magnitude on a short timescale of several years (e.g., \citealt{2015ApJ...800..144L}; \citealt{2016MNRAS.457..389M}; \citealt{2016ApJ...826..188R}). The AGNs that have shown this type of event are collectively  referred to as ``changing-look AGNs (CLAGNs)''. Hereafter, this type of transition phenomenon and its process are referred to as a ``changing-look phenomenon'' and  ``changing-look process,'' respectively.

The following three  hypotheses have been so far proposed to explain the physical mechanisms that cause changing-look phenomena.  
One  explains it in terms of the motion of obscuring clouds into or out of the line of sight, as  supported by soft X-ray  observations (e.g., \citealt{2003MNRAS.342..422M}). 
One  assumes a tidal disruption event (TDE) occurring in an AGN (e.g., \citealt{2015MNRAS.452...69M}). 
The other  proposes that a  drastic change in the mass accretion rate  causes a drastic change in the luminosity and the spectral energy distribution (SED) of the AGN, which strongly affects BLR materials (e.g., \citealt{2018MNRAS.480.3898N}). 
 Since the first  one is not caused  due to intrinsic changes in the central engine and since the second  one seems  very rare, we hereafter mainly focus on the third  one as  the intrinsic changing-look phenomenon.

\cite{2018MNRAS.480.3898N} studied SED variations through a changing-look process from type-1 (bright)  to type-1.9 (faint)  states on a typical intrinsic CLAGN, Mrk~1018. 
They  found that the disk black-body emission from a standard disk (\citealt{1984PASJ...36..741M}; \citealt{1986ApJ...308..635M}) and/or warm Comptonization (soft X-ray excess) emission from a warm disk (e.g., \citealt{2011PASJ...63S.925N}; \citealt{2013PASJ...65....4N}) were dominant during the type-1 phase,  whereas the hot Comptonization continuum from a hot corona (hot accretion flow; e.g., \citealt{2014ARA&A..52..529Y}) was dominant  during the type-1.9 phase. 
In addition, they found that the SED dramatically changed at the transition  with  an Eddington ratio $L/L_{\rm Edd}$ of  a few percent. 
 Since the SED change and the transition point are consistent with those of the high/soft-to-low/hard state transition in Galactic black hole binaries, they suggested that the intrinsic changing-look processes were caused by the state transition between the high/soft and low/hard states of an accretion flow onto a SMBH. 
 Since then,  their hypothesis has been verified  in other CLAGNs (e.g., \citealt{2019ApJ...883...76R}; \citealt{2020MNRAS.491.4925G}). 
However, it is still unclear how the mass-accretion rate changes  drastically in such a short period, which is orders of magnitude shorter than the viscous timescale of  the standard disk. 
The necessity of further studies of the disk structure and the disk instability  in the vicinity of a SMBH  has been recognized  among astronomers in the field. 

Another  critical question on intrinsic CLAGNs is  what happens in BLR materials  before and after the drastic luminosity and SED variations of the emission from the accretion flow onto a SMBH. 
 Given that optical broad emission-lines appear or disappear  before/after  a changing-look process, one possible and perhaps most intuitive scenario is that BLR materials appear in the type-1 phase and disappear in the type-2 phase. 
 Alternatively, it is also possible that the BLR materials remain at the same location in either state and  are activated  enough to produce optical broad lines in the type-1 phase,  whereas they are deactivated not to  emit optical broad-lines anymore in the type-2 phase. 
   Conventional optical spectroscopic observations are unsuitable for distinguishing these two scenarios because AGNs do not emit optical broad lines  during the type-2 phase by definition. 
Therefore, another approach is required to address  the question. 

In order to constrain what happens in  BLR materials through the changing-look process, we  here study emission in  the X-ray band, especially a narrow Fe-K$\alpha$ emission line at 6.4~keV, from a CLAGN. 
An Fe-K$\alpha$ line is  generated  through  photo-absorption to a hard X-ray continuum originating from a corona near a SMBH caused by  materials in the accretion disk, BLR, dusty torus and/or  distant molecular disk/clouds. 
Note that we use the term ``narrow Fe-K$\alpha$ line''  to mean a non-relativistic iron line with  a velocity width from zero  to several thousands~km~s$^{-1}$,  which encompasses the  velocity widths  of both broad and narrow emission lines in the optical band (see \S\ref{4.1.2}). 

As  the target CLAGN, we select NGC~3516, which  used to be recognized as a bright and typical type-1 Seyfert with strong broad emission lines but around 2014 got into the faint type-1.9 or -2 phase, at which the broad lines almost disappeared \citep{2019MNRAS.485.4790S}. 
\cite{2022ApJ...925...84M} analyzed the X-ray data of NGC~3516 derived during the faint phase at 2017, and concluded that the changing-look process can be explained by the change of the ionizing SED without requiring any new or variable obscuration.   
Therefore, NGC~3516 is an intrinsic CLAGN.  
In 2020, it returned to the type-1 phase, where optical broad lines reappeared \citep{2021MNRAS.505.1029O}.
We performed X-ray and optical simultaneous monitoring observations  between 2013  and 2014, during which NGC~3516 was almost in the type-2 phase, with the X-ray astronomical  satellite \textit{Suzaku} and  optical 1.5-m class ground telescopes in Japan \citep{2016ApJ...828...78N}. 
In the present study, we investigate the variability of the narrow Fe-K$\alpha$ emission line at 6.4~keV on the timescale of a week to a year which corresponds to radii from a BLR to a dusty torus, and conduct the ``narrow Fe-K$\alpha$ reverberation mapping'' by comparing it  with the X-ray and optical continuum variabilities. 

\renewcommand{\arraystretch}{1}
\begin{table*}[t]
\caption{X-ray observations in the faint phase of NGC~3516 by \textit{Suzaku}.}
 \begin{center}
 \begin{tabular}{cccccc}
\hline
Epoch &
ObsID&
Observation Start (UT)&
Observation End (UT)&
Middle (MJD)&
Exposure (ksec) \\\hline

1 & 
708006010 &
2013-04-09T23:13:20 &
2013-04-11T01:06:16 &
56392.5 &
51.40 \\

2 &
708006020 &
2013-04-27T00:17:13 &
2013-04-27T10:42:22 &
56409.2 &
19.12 \\

3 &
708006030 &
2013-05-12T00:22:24 &
2013-05-13T02:30:23 &
56424.6 &
50.43 \\

4 &
708006070 &
2013-05-23T03:32:08 &
2013-05-24T07:05:07 &
56435.7 &
51.48 \\

5 &
708006040 &
2013-05-29T11:02:50 &
2013-05-30T15:15:14 &
56442.0 &
54.22 \\

6 &
708006060 &
2013-11-04T06:15:13 &
2013-11-05T05:10:17 &
56600.7 &
46.20 \\

7&
708006080 &
2014-04-07T16:54:26 &
2014-04-08T12:00:24 &
56755.1 &
51.54 \\

8 &
710009010 &
2015-05-12T08:01:41 &
2015-05-15T00:50:12 &
57155.7 &
117.7 \\ \hline\hline

\end{tabular}
\end{center}

\label{tab:suzaku_obs}
\end{table*}
%%%%%%%%%%%Table 1%%%%%%%%%%%% 

The present paper is organized as follows.  
Observations and data reductions are summarized in section~\ref{two} (n.b., part of the data  were described in \citealt{2016ApJ...828...78N}). 
Spectral analyses  of the X-ray continuum and narrow Fe-K$\alpha$ line, light-curves and timing analyses, including  cross-correlation and JAVELIN analyses, are described in section~\ref{three}. 
Finally, we discuss what happens in  the BLR materials before and after transitions and derive constraints to the BLR origin  in section~\ref{four},  comparing our results with  previous multi-wavelength studies  and those of the narrow Fe-K$\alpha$ line. 
We adopt  cosmological parameters of $H_0 = 73$~km~s$^{-1}$~Mpc$^{-1}$, $\Omega_{\Lambda} = 0.73$ and $\Omega_{\rm m} = 0.27$ throughout the present paper. 
Errors quoted in this paper refer to 1$\sigma$ errors unless noted otherwise.

%=================
\section{Observations and Data Reduction}
\label{two}
%=================

%\renewcommand{\arraystretch}{1}
%\begin{table*}[t]
%\caption{Ground-based optical telescopes and instruments of the $B$-band observations.}
% \begin{center}
% \begin{tabular}{cccccc}
%\hline
%Telescope &
%Instrument &
%Mirror diameter (m)&
%Field of view (arcmin)&
%Pixel scale (arcsec pixel$^{-1}$)&
%Number of observations \\\hline
%
%Pirka &
%1.6 &
%MSI$^{a}$ &
%$3.3 \times 3.3$ &
%0.39 &
%86 \\
%
%Kiso Schmidt &
%1.5 &
%KWFC$^{b}$ &
%$60 \times 30^{c}$ &
%0.95 &
%32  \\
%
%Nayuta &
%2.0 &
%MINT$^{d}$ &
%$11 \times 11$ &
%0.32 &
%31 \\
%
%Kanata &
%1.5 &
%HOWPol$^{e}$ &
%$\phi~15$ &
%$0.29$ &
%31 \\\hline\hline
%
%
%\end{tabular}
%\end{center}
%\tablenotetext{a}{Multi-Spectral Imager (MSI; Watanabe et al. 2012).}
%\tablenotetext{b}{Kiso Wide Field Camera (KWFC; Sako et al. 2012).}
%\tablenotetext{c}{Only one of the eight CCDs installed in KWFC was used.}
%\tablenotetext{d}{Multiband Imager for Nayuta Telescope (MINT; Ozaki et al. 2005).}
%\tablenotetext{e}{Hiroshima One-shot Wide-field Polarimeter (HOWPol; Kawabata et al. 2008).}
%\label{tab:optical_obs}
%\end{table*}
%%%%%%%%%%%%Table 1%%%%%%%%%%%% 

%%%%%%%%%%%Table 1%%%%%%%%%%                                                                                                                              
\renewcommand{\arraystretch}{1}
\begin{table*}[t]
\caption{Results of the spectral fits to the eight {\it Suzaku} datasets of NGC~3516.  }
 \begin{center}
 \begin{tabular}{lccccccccc}
\hline
&
MJD &
56392.5&
56409.2&
56424.6&
56435.7&
56442.0&
56600.7&
56755.1&
57155.7
\\\hline
Model &
  Parameter &
  Epoch 1 &
  Epoch 2 &
  Epoch 3 &
  Epoch 4 &
  Epoch 5 &
  Epoch 6 &
  Epoch 7 &
  Epoch 8 \\ \hline
\texttt{phabs} &
  $N_{\rm H}^{a}$ &
  $1.35_{-0.22}^{+0.23}$ &
  $1.49\pm 0.18$ &
  $0.91 \pm 0.09 $ &
  $1.03 \pm 0.08$ &
  $0.93\pm 0.10$ &
  $1.31_{-0.28}^{+0.29}$ &
  $2.93_{-0.39}^{+0.40}$ &
  $1.57 \pm0.07$ \\[1.5ex]
 \texttt{zashift} &
 $z$ &
  \multicolumn{8}{c}{0.00884 (fix)} \\[1.5ex]

\texttt{powerlaw} &
  $\Gamma$ &
  $1.34 \pm 0.06$ &
  $1.61 \pm 0.05$ &
  $1.63 \pm 0.03$ &
  $1.75\pm 0.02$ &
  $1.62\pm 0.03$ &
  $1.45 \pm 0.08$ &
  $1.49_{-0.10}^{+0.11}$ &
  $1.60 \pm 0.02$ \\
 &
  $F_{\rm 2-10}^{b}$ &
  $0.44 \pm 0.01$ &
  $1.52\pm 0.02$ &
  $2.14\pm 0.02$ &
  $2.68\pm 0.02$ &
  $1.60\pm 0.01$ &
  $0.31 \pm 0.01$ &
  $0.21 \pm 0.01$ &
  $1.64 \pm 0.01$ \\[1.5ex]
  
\texttt{gaussian1} &
  $E$ &
  \multicolumn{8}{c}{6.4 keV (fix)} \\
 &
  $F_{\rm 6.4}^{c}$&
  $2.37\pm 0.15$ &
  $2.03\pm 0.32$ &
  $3.18_{-0.23}^{+0.24}$ &
  $3.47\pm 0.25$ &
  $3.42\pm 0.21$ &
  $1.8\pm 0.14$ &
  $1.57\pm 0.12$ &
  $2.78\pm 0.15$ \\
  
   &
   EW$_{\rm 6.4}^{d}$&
  $459.5^{+34.7}_{-27.1} $ &
  $120.5^{+20.6}_{-19.6} $ &
  $134.9^{+10.3}_{-9.4}$ &
  $121.5^{+9.4}_{-8.6}$ &
  $192.8^{+12.3}_{-11.6}$ &
  $ 506.2^{+50.1}_{-36.5}$ &
  $ 667.9^{+70.6}_{-46.9}$  &
  $ 151.6^{+8.8}_{-8.1}$  \\[1.5ex]
  
\texttt{gaussian2} &
  $E$ &
  \multicolumn{8}{c}{6.97~keV (fix)} \\
 &
  $F_{\rm 6.97}^{e}$ &
  $4.8\pm 1.2$ &
  $4.4 \pm 3.1$ &
  $<4.1$ &
  $3.5\pm 2.4$ &
  $3.3\pm 1.9$ &
  $1.9 \pm 1.0$ &
  $4.2\pm 1.0$ &
  $3.0\pm 1.4$ \\
     &
   EW$_{\rm 6.97}^{f}$&
  $95.6^{+26.1}_{-21.9} $ &
  $27.9^{+19.8}_{-18.6} $ &
  $< 18.2$ &
  $13.0^{+8.5}_{-8.4}$ &
  $19.8^{+11.4}_{-10.4}$ &
  $56.2^{+31.7}_{-30.2}$ &
  $188.9^{+55.3}_{-43.0}$ &
  $17.4^{+8.5}_{-8.1}$ \\ [1.5ex]
  
   &
   $L_{\rm 2-10}^{g}$&
  $0.72 \pm 0.02$ &
  $2.33 \pm 0.07 $ &
  $3.43 \pm 0.03$ &
  $4.23 \pm 0.03$ &
  $2.58 \pm 0.02$ &
  $0.51 \pm 0.02 $&
  $0.32 \pm 0.02$ &
  $2.52 \pm 0.02$ \\ [1.5ex]
 
 \hline
$\chi^2$ / d.o.f. &
  \multicolumn{1}{l}{} &
  $456.2/517$ &
  $552.5/598$ &
  $1446.5/1503$ &
  $1570.7/1579$ &
  $1434.5/1393$ &
  $339.6/356$ &
  $210.9/261$ &
  $1884.0/1731$ \\ \hline

\end{tabular}
\end{center}
\tablenotetext{a}{Equivalent hydrogen column density in  $10^{22}$ cm$^{-2}$.}
\tablenotetext{b}{ Unabsorbed 2--10 keV flux of the \texttt{powerlaw} component in  $10^{-11}$~erg~s$^{-1}$~cm$^{-2}$.}
\tablenotetext{c}{ Unabsorbed flux of \texttt{gaussian1} at 6.4~keV  in  $10^{-13}$~erg~s$^{-1}$~cm$^{-2}$.}
\tablenotetext{d}{Equivalent width of \texttt{gaussian1} at 6.4~keV in  eV.}
\tablenotetext{e}{ Unabsorbed flux of \texttt{gaussian2} at 6.97~keV  in  $10^{-14}$~erg~s$^{-1}$~cm$^{-2}$.}
\tablenotetext{f}{Equivalent width of \texttt{gaussian2} at 6.97~keV in  eV.}
\tablenotetext{g}{The 2--10 keV luminosity in  $10^{42}$~erg~s$^{-1}$.}
\vspace{0.5cm}
\label{tab:suzaku}

\end{table*}
%%%%%%%%%%%Table 1%%%%%%%%%%%% 

We performed X-ray and optical simultaneous monitoring  of the CLAGN NGC~3516 from 2013 April to 2014 April, during which NGC~3516 was  in the type-2 (faint) phase (PI: H. Noda). 
The X-ray observations were performed on seven  occasions with the X-ray astronomical satellite \textit{Suzaku}. 
The first five (Epochs~1--5) were conducted with intervals of 1--2 weeks, and the last two (Epochs~6--7) were with intervals of about half a year, all with exposures of $\sim 50$~ksec except for the  second observation with $\sim19$~ksec.
The details of the observations are same as those of \cite{2016ApJ...828...78N}, and presented in Table~\ref{tab:suzaku_obs}.
The data of the X-ray Imaging Spectrometer (XIS; \citealt{2007PASJ...59S..23K}) were obtained with its normal mode and processed  with the software version 2.4. 
The front-illuminated XIS0 and 3 data were merged and used  for our analyses,  whereas the back-illuminated XIS1 data were not used because  the background around 6~keV was high. 
We extracted source events from a  circular region with  a radius of $180"$ and background events  from an annular region with  inner and outer radii of $270"$ and  $360"$, respectively. 
The response and ancillary-response file were produced  with the software packages \textit{xisrmfgen} and \textit{xissimarfgen}
\citep{2007PASJ...59S.113I}, respectively, in HEASOFT-6.28. 
In the present paper, we did not use the data of the Hard X-ray Detector (HXD; \citealt{2007PASJ...59S..35T}). 

In addition to the data of the densely-sampled seven \textit{Suzaku} observations, we used more X-ray data as  follows. 
We observed NGC~3516  with \textit{Suzaku} one more time on 2015 May 12 (Epoch~8) with an exposure time of $\sim118$~ksec
as shown in Table~\ref{tab:suzaku_obs}. 
We reduced the data  in the same way as the previous seven \textit{Suzaku} datasets and  combined  them in our X-ray analyses. 
Furthermore, the Neil Gehrels \textit{Swift} Observatory (hereafter, \textit{Swift}) monitored NGC~3516 in 2012--2013 and 2014--2015, the timings of which were shortly before and after  our intensive X-ray--optical simultaneous monitoring in 2013--2014, respectively.
We used the data of the \textit{Swift} X-Ray Telescope (XRT) with the Photon Counting (PC) mode in X-ray analyses together with \textit{Suzaku}/XIS.
The number of the XRT datasets with the PC mode is 56. 
We  reduced the XRT data with the automated pipeline  provided by the UK \textit{Swift} Science Data  Centre (\citealt{2007A&A...469..379E}; \citealt{2009MNRAS.397.1177E}) and obtained X-ray spectra. 
In the present paper, we did not use the Ultraviolet/Optical Telescope (UVOT) and Burst Alert Telescope (BAT) data of the \textit{Swift} observations. 

Optical observations of NGC~3516 were performed in the $B$ band  with  four ground-based telescopes in Japan, Pirka, Kiso Schmidt, Nayuta, and Kanata. 
The \textit{Suzaku} observations in Epochs~1--7 were simultaneously covered, 
and in addition, the intervals between them were also densely covered, typically once per  day. 
The image data were reduced  in the standard manner for CCD detectors  with IRAF.
We derived the $B$-band flux of  NGC~3516  in every image with  differential image photometry, 
in which we subtracted a reference image from every image by matching the point-spread function profile 
and performed  aperture photometry using reference stars.
We  successfully derived a $B$-band variable component after subtracting a stable component including the host galaxy emission.
The detail of  the method is described in  \cite{2016ApJ...828...78N}.
The flux-density data of the $B$-band light curve derived  with the differential image photometry were published in \cite{2016yCat..18280078N}. 
  In the present analyses, we used flux density values  with an artificial 0.5~mJy offset in order to circumvent negative flux densities, which happened, depending on the choice of the reference image (see Fig.~4 in \citealt{2016ApJ...828...78N}). The offset does not affect the timing analyses in \S\ref{three}.

%%%%%%%%%%%%%%%%figure1%%%%%%%%%%%%%%%%%%%%%%%%
\begin{figure}[t]
\epsscale{1.2}
\plotone{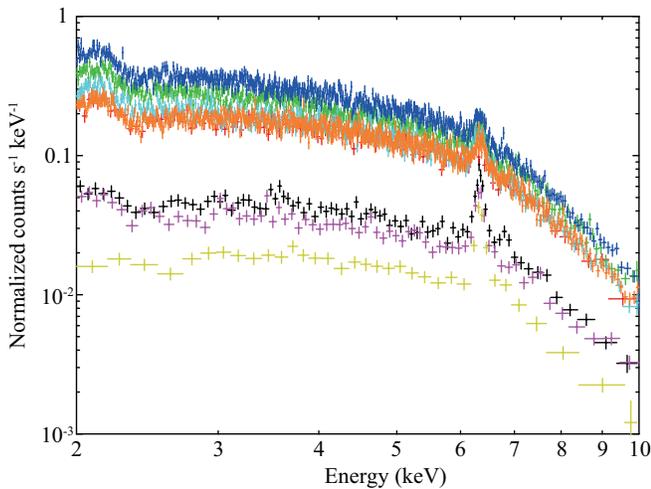}
\caption{ X-ray spectra of NGC~3516 obtained  in the eight \textit{Suzaku} observations in black, red, green, blue, cyan, magenta, yellow, and orange  for Epochs~1 to 8, respectively.}
\label{suzaku_spec}
\end{figure}
%%%%%%%%%%%%%%%%figure1%%%%%%%%%%%%%%%%%%%%%%%%

%%%%%%%%%%%%%%%%figure2%%%%%%%%%%%%%%%%%%%%%%%%
\begin{figure*}[t]
\epsscale{1.1}
\plotone{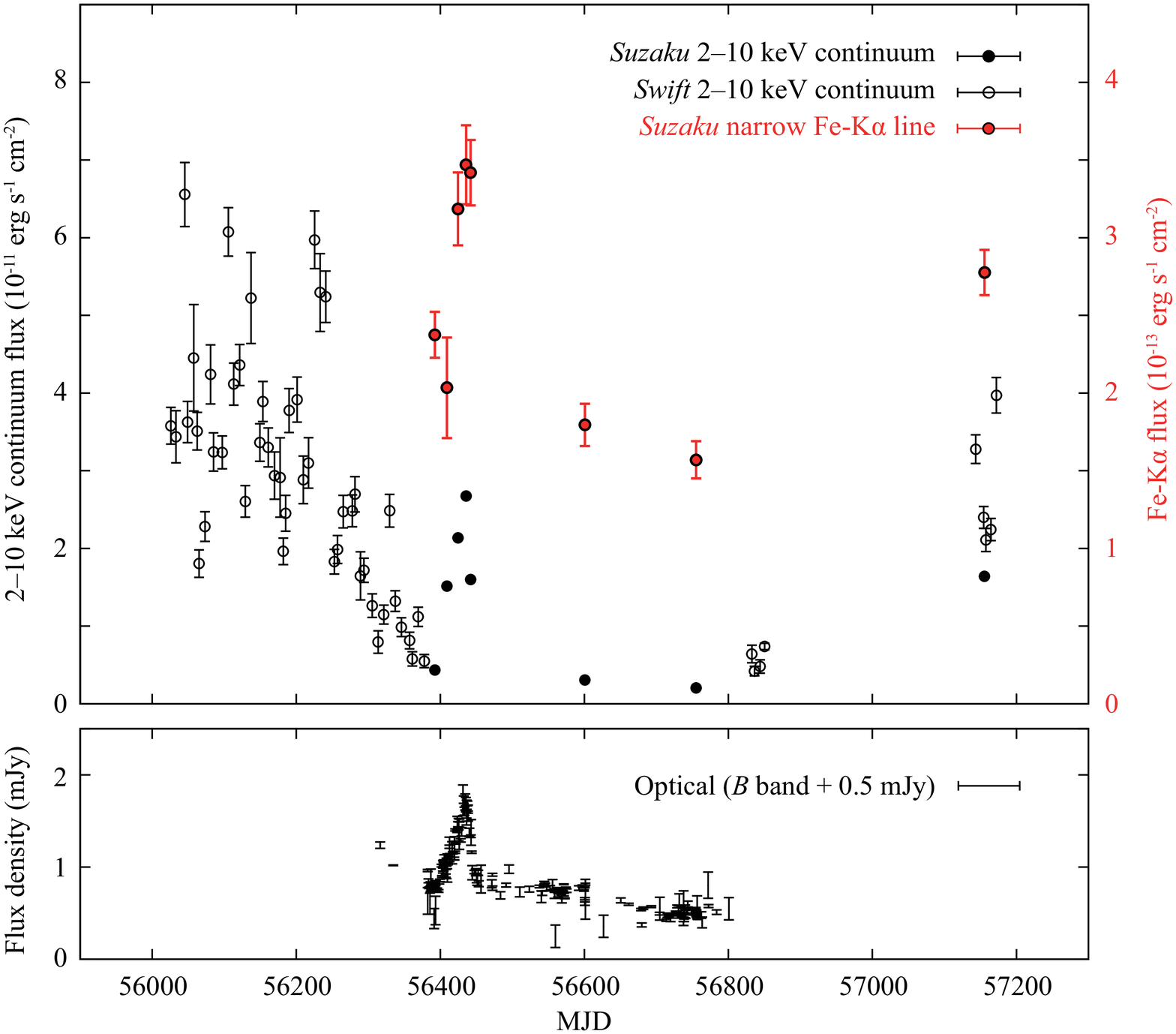}
\caption{(Top)  2--10~keV continuum flux in black and narrow Fe-K$\alpha$ emission line flux in red. Filled and open circles are  the data of \textit{Suzaku} (Table~\ref{tab:suzaku}) and \textit{Swift}, respectively. (bottom)  $B$-band light curve with an 0.5~mJy offset. }
\label{LCs}
\end{figure*}
%%%%%%%%%%%%%%%%figure2%%%%%%%%%%%%%%%%%%%%%%%%

%==================================================
\section{Data Analyses and Results}
\label{three}
%==================================================
%==================================================
\subsection{X-ray, optical, and narrow Fe-K$\alpha$ light curves}
\label{3.1}
%==================================================

Figure~\ref{suzaku_spec} showed  the extracted 2--10~keV spectra from the \textit{Suzaku} data. 
The spectra in Epochs~1--7 are  practically the same as those in Figure~3 in \cite{2016ApJ...828...78N} because the datasets are same,  whereas the spectrum in  Epoch~8 (orange in Figure~\ref{suzaku_spec}) was newly added in this study. 
We fitted them individually with  a model of a photo-absorbed power-law function and two Gaussians,  or \texttt{phabs*zashift*(cflux*powerlaw + cflux*gaussian + cflux*gaussian)} in the XSPEC notation. 
For convenience, we hereafter express the first and second \texttt{gaussian} models as \texttt{gaussian1} and \texttt{gaussian2}, respectively. 
Here, the power-law model reproduces an X-ray continuum from a corona, 
\texttt{gaussian1} models a narrow and neutral Fe-K$\alpha$ emission line at $6.4$~keV, and \texttt{gaussian2} models a narrow hydrogen-like Fe emission-line at 6.97~keV. 
The use of \texttt{gaussian1} is mainly different from the model employed in  \cite{2016ApJ...828...78N}, where they used the \texttt{pexmon} model \citep{2007MNRAS.382..194N}. 
The \texttt{pexmon} model  includes a reflection continuum in addition to the narrow Fe-K$\alpha$ line  as in   \texttt{gaussian1}.
However, the \texttt{pexmon} flux was  determined almost  solely by the narrow Fe-K$\alpha$ line  because the reflection component is much weaker than  the power-law continuum in this band, 2--10 keV. 
Hence, we  adopted \texttt{gaussian1} in this study for simplicity,  given that the difference of the models  would make no significant difference in the results. 
In the fitted model, we considered the redshift ($z$) by convolving \texttt{zashift} with $z$ fixed at 0.00884. 
We convolved the \texttt{cflux} models to the individual components to measure their fluxes.
In the spectral  fitting, the column density ($N_{\rm H}$) of \texttt{phabs} and the photon index ($\Gamma$) of \texttt{powerlaw} were  allowed to vary. 
The energies of the two \texttt{gaussian} components were fixed at 6.4~keV and 6.97~keV,  and so were their sigmas at 0~eV. 
The fluxes of all the \texttt{cflux} models were  allowed to vary, and their minimum and maximum energies between which the fluxes  were calculated were fixed at 2 and 10~keV, respectively. 
Spectral fitting was based on  the $\chi^2$ statistics. 
As a result, all the fits were  broadly successful with reduced $\chi^2 < 1.09$. Table~\ref{tab:suzaku} tabulates the fitting result for all   the eight \textit{Suzaku} observations  of NGC~3516.

To check the presence of a broader component around 6.4~keV, we added one more gaussian model with the energy fixed at 6.4~keV, and the sigma and normalization left free. 
To also check the energy shift of \texttt{gaussian1}, we allowed its energy to vary, and fitted all the \textit{Suzaku} spectra. 
As a result, a broad component with the sigma consistent with 700~eV within $1\sigma$ errors was not ruled out in all the epochs except for Epoch~7 in which the source was too faint to constrain the broad component. 
The energy and flux of \texttt{gaussian1}, and the 2--10~keV flux of \texttt{powerlaw} became consistent within $1\sigma$ errors with those in the fits without the broad component (Table~\ref{tab:suzaku}). 
Therefore, the analyses from \S\ref{3.2} on were not affected by ignoring the presence of the broad component.   
As modeled by \texttt{pexmon} in \cite{2016ApJ...828...78N}, most of the broad component is considered to be a part of a non-relativistic reflection continuum with the Fe-K absorption edge at 7.1~keV because it is accompanied by the narrow Fe-K$\alpha$ line, although a relativistically-blurred reflection and/or a partially-absorbed power-law continuum might be included. 
We also fitted all the \textit{Suzaku} spectra allowing the energy of \texttt{gaussian2} to vary, and confirmed that it can be located in a range of $6.7$--7.2~keV considering $1\sigma$ errors. 
This range includes energies of not only hydrogen-like Fe but also helium-like Fe and Fe-K$\beta$ emission lines.
Whereas the Fe-K$\beta$ line is associated with the Fe-K$\alpha$ line at 6.4~keV, the hydrogen- and helium-like Fe lines must be emitted in a highly ionized region, which might be a radiatively inefficient accretion flow as reported in low-luminosity AGNs (e.g., \citealt{2004ApJ...607..788D}). 
The detailed interpretations of the broad component and the highly-ionized emission lines are beyond the scope of this paper, they will be discussed in subsequent papers.

%%%%%%%%%%%%%%%%figure3%%%%%%%%%%%%%%%%%%%%%%%%
\begin{figure*}[t]
\epsscale{1}
\plotone{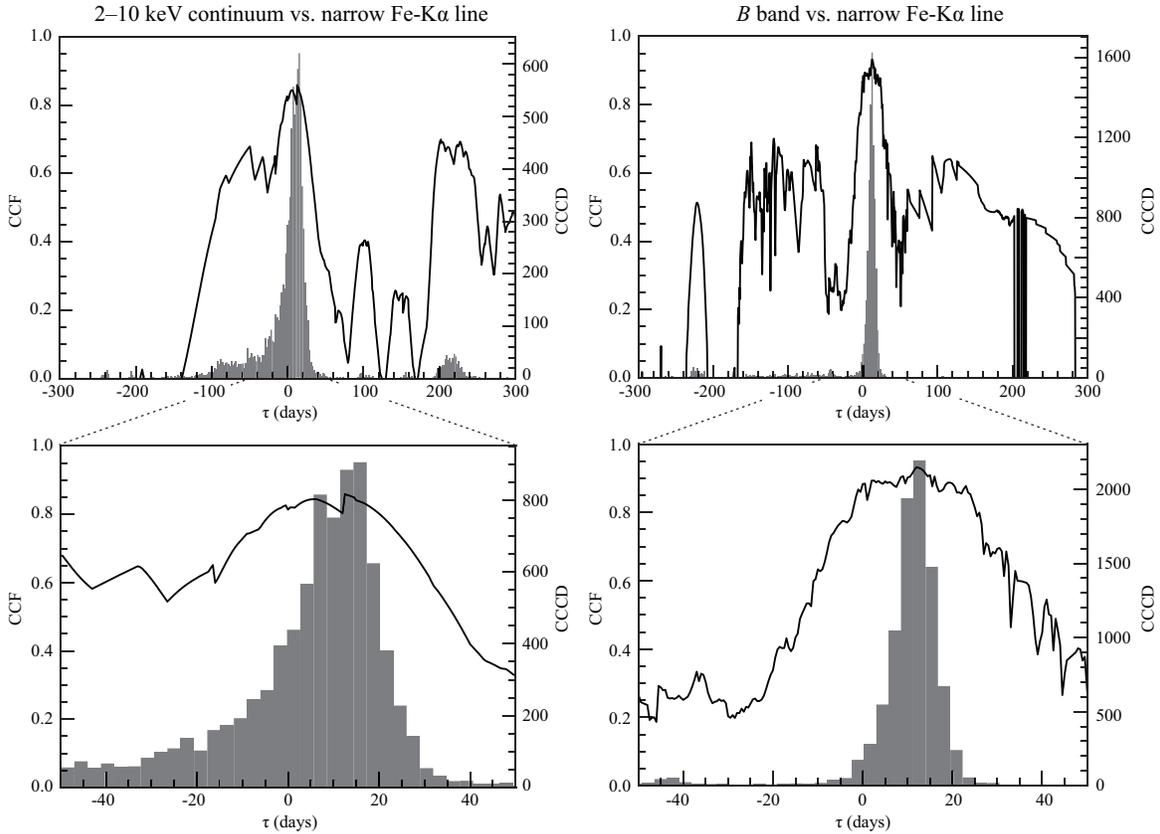}
\caption{Left panels show the results of ICCF analyses  of the NGC~3516 data for the time lag $\tau$ of $-300$ to $300$~days  between the 2--10~keV continuum and the narrow Fe-K$\alpha$ light curves. A positive value of $\tau$ indicates that 2--10~keV continuum flux precedes the corresponding narrow Fe-K$\alpha$ flux. Right panels show those  with the $B$-band and narrow Fe-K$\alpha$ light curves. Bottom panels show zoomed-up histograms  for a range of $-50$ to $50$~days in the top panels. 
Black solid lines show cross-correlation functions (CCFs),  and gray histograms show the distribution of the CCF centroids (CCCDs).}
\label{CCF}
\end{figure*}
%%%%%%%%%%%%%%%%figure3%%%%%%%%%%%%%%%%%%%%%%%%

We analyzed the \textit{Swift} data as follows.  
We first made an averaged spectrum in 2--10~keV  from  56 datasets and fitted it with  a model of \texttt{phabs*zashift*(cflux*powerlaw + cflux*gaussian1)}  with $\chi^2$ statistics. 
Here, we did not include  \texttt{gaussian2} in the model, unlike our analysis of the \textit{Suzaku} data, because of low statistics in the \textit{Swift} spectra at energies around $\sim 6.97$~keV. 
The energy and sigma of \texttt{gaussian1} were fixed at 6.4~keV and 0~eV, respectively. 
The parameters $N_{\rm H}$ of \texttt{phabs}, $\Gamma$ and flux of \texttt{powerlaw}, and the flux of \texttt{gaussian1} were  allowed to vary.
The redshift  in \texttt{zashift} was fixed at  $z=0.00884$.  
The resultant fit  was successful with $\chi^2/$d.o.f.$=295.14/289$. 
The best-fit \texttt{gaussian1} flux  was $(3.78 \pm 1.02) \times 10^{-12}$~erg~s$^{-1}$~cm$^{-2}$ ($1\sigma$ error). 
Even in the averaged spectral fit merging 56 data, the uncertainty of the narrow Fe-K$\alpha$ flux is larger than that of a \textit{Suzaku} single epoch by a factor of $\gtrsim 4$ because of low sensitivity.
Hence, in the following analysis, we did not use the \textit{Swift} data to constrain the narrow Fe-K$\alpha$ flux.  
Then, we  extracted a 2--7~keV spectrum from every \textit{Swift} data,  totaling 56 spectra, and fitted the individual spectra with basically the same model but with the C-statistics. 
Here,  we ignored the 7--10~keV band in individual spectra because counts  at energies above 7~keV are too low. 
In the fitting, we  fixed $N_{\rm H}$ of \texttt{phabs} and $\Gamma$ of \texttt{powerlaw} at the values obtained in the averaged-spectral fit.  
Again, the narrow Fe-K$\alpha$ fluxes  could not be constrained in the individual \textit{Swift} spectra because of too low statistics. 
As a results, all the fits were acceptable, and we determined the 2--10~keV fluxes from the individual \textit{Swift} spectra. 

Figure~\ref{LCs} shows the 2--10~keV continuum light-curves for $\sim$1200~days obtained  from the \textit{Suzaku} and \textit{Swift} spectral analyses. 
The 2--10~keV continuum flux almost monotonously decreased from MJD~56000 to $56390$ and got into the faint phase. 
Then, it showed a significant flare by an order of magnitude from MJD~56393 to 56440 peaking at MJD~56436, and returned into the faint phase. 
The faint phase continued until MJD~56800, and it re-brightened around MJD~57200. 
Figure~\ref{LCs} also shows the $B$-band light curve,  which has similar variation to the 2--10~keV light curve.   
As already reported by \cite{2016ApJ...828...78N}, the 2--10~keV continuum and the $B$-band light curves showed a  strong positive correlation with a high cross-correlation coefficient of $>0.95$. 
\cite{2016ApJ...828...78N} performed the Interpolation Cross Correlation Function (ICCF) and  JAVELIN analyses (see \S\ref{3.2} for details of the methods) and  determined the $B$-band time lag behind the X-ray continuum to be $2.0^{+0.7}_{-0.6}$~days ($1\sigma$ errors).
 
 %%%%%%%%%%%%%%%%figure4%%%%%%%%%%%%%%%%%%%%%%%%
\begin{figure*}[t]
\epsscale{1.2}
\plotone{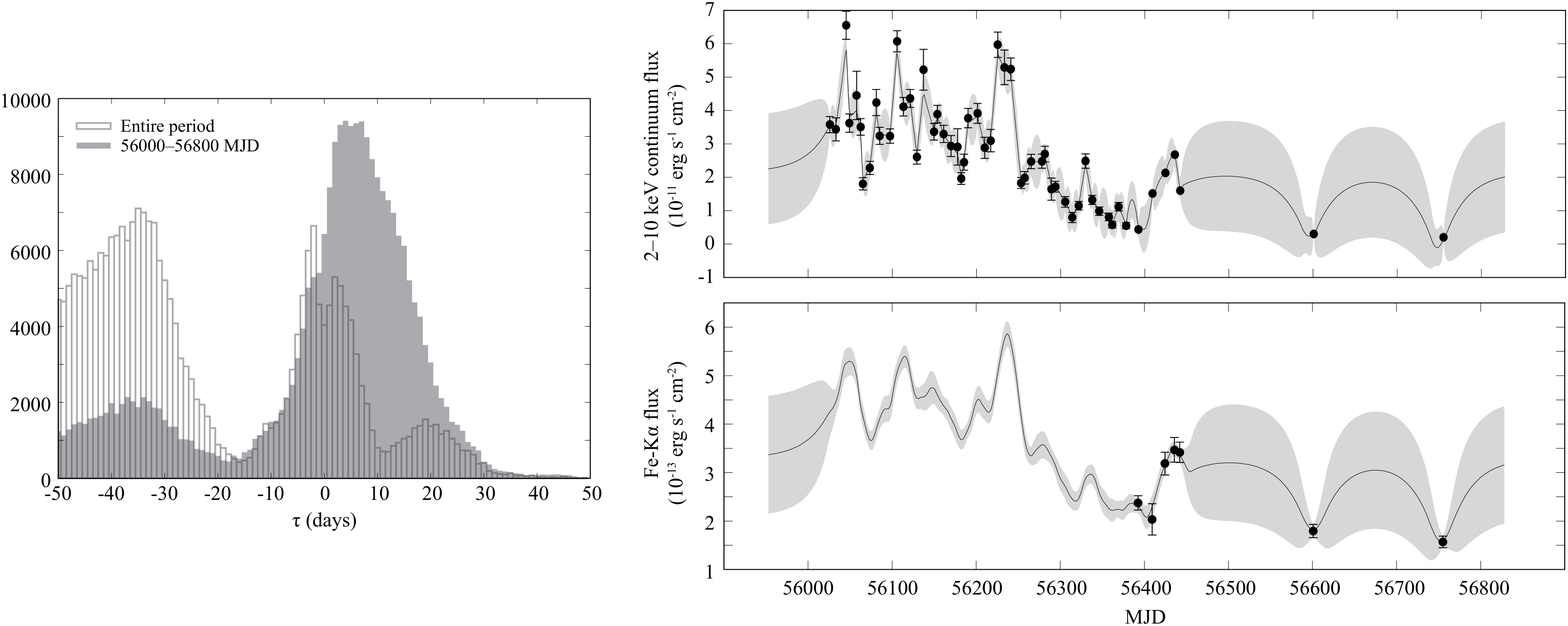}
\caption{Results of the JAVELIN analyses,  using the 2--10~keV continuum and narrow Fe-K$\alpha$ light-curves as the primary and response light-curves, respectively. Left panel shows the posterior distributions of the time lag $\tau$ obtained  for the (open boxes) entire period  and (filled boxes) MJD~56000--56800. Right panels show the best-fit light-curve models of the (top panel) 2--10~keV continuum flux and  (bottom) narrow Fe-K$\alpha$ flux. The shaded regions correspond to $1\sigma$ uncertainties. }
\label{JAVELIN_suzaku}
\end{figure*}
%%%%%%%%%%%%%%%%figure4%%%%%%%%%%%%%%%%%%%%%%%%
 
Figure~\ref{LCs} also shows the narrow Fe-K$\alpha$ line light curve. 
The narrow Fe-K$\alpha$ flux  also showed  an apparent flare during MJD~56409--56442 with its peak likely follow those of the 2--10~keV  and $B$-band light curves. 
After the flare, it became faint until MJD~56755 and re-brightened at MJD~57156, similarly to the 2--10~keV light curve.
We fitted the narrow Fe-K$\alpha$ light curve from MJD~56393 to 57156 (Epochs~1--8) with a constant model and found that 
 the  model was rejected with $\chi^2/$d.o.f. $= 111/6$,  proving that the  observed time-variations  are significant. 
 Fitting only five points in the peak at MJD~56393--56442 (Epochs~1--5)  with a constant model  resulted in $\chi^2/$d.o.f. $= 15.1/3$, confirming the variability. 
Therefore, we  conclude that  the narrow Fe-K$\alpha$ flux of NGC~3516 showed a significant variation on  a timescale within $\sim 50$~days in MJD~56393--56442. 

Table~\ref{tab:suzaku} shows that the equivalent width (EW) of the narrow Fe-K$\alpha$ line reached $\sim 670$~eV (Epoch~7), which is larger by a factor of $\sim 7$ than those in typical type-1 AGNs (e.g., \citealt{2013MNRAS.435.1840R}). 
 The main reason for the large EW appears to be not an intrinsic increase of the line flux but a decrease of the continuum flux; while the 2--10~keV continuum flux  decreased by a factor of $\sim 13$, the narrow Fe-K$\alpha$ flux varied only by a factor of $\sim 2$. 
This indicates that the narrow Fe-K$\alpha$ flux consists of a variable component  that  varied on  timescales of a week to months and a stationary component  that remained stable for a year (Epochs~1--7).
Here, we can regard the lowest Fe-K$\alpha$ flux of $\sim 1.6 \times 10^{-13}$~erg~s$^{-1}$~cm$^{-2}$  at Epoch~7 as the upper limit of the stationary flux during our observation period, because the variable 2--10~keV continuum flux was minimum in  the epoch.  
We discuss the variable and stationary fluxes furthermore in \S\ref{3.4}.

%==================================================
\subsection{ICCF analyses}
\label{3.2}
%==================================================

 Using the high-quality long-term 2--10~keV continuum, $B$-band, and narrow Fe-K$\alpha$ light curves,  we  determine the amount of the time lag ($\tau$) of the narrow Fe-K$\alpha$ flux  variation to the X-ray and/or optical continuum variations. 
We started from the standard cross-correlation analysis called ICCF \citep{1998PASP..110..660P} in this subsection, which has been employed in many AGN reverberation studies, before employing the more complex JAVELIN method in the next subsection. 
 The ICCF method interpolates  neighboring two data points  with  a linear function and conducts the traditional cross-correlation analysis. 
 The method can be applied to two light curves  regardless of whether they were  simultaneously sampled or not.  
In the present paper, we employed the \texttt{PYCCF} algorithm\footnote{We employed the code \texttt{pyCCF} in http://ascl.net/code/v/1868}. 

We first applied the ICCF analysis  to a pair of the 2--10~keV continuum and narrow Fe-K$\alpha$ light curves (Figure~\ref{LCs}~(top)) and estimated their time lag $\tau$.  
The range of $\tau$ was set from $-300$ to $300$~days. 
The number of iterations was  $10000$. 
The threshold of the $r$ value to judge correlations significant was set to $0.4$, which corresponds to the $99$\% significance level with d.o.f. $= 40$. 
In error estimation, we employed the Flux Randomization (FR) and Random Subset Selection (RSS). 
Figure~\ref{CCF} left-panels show the derived cross-correlation functions (CCF) and the distribution of the CCF centroid (CCCD) of $\tau$.     
The time lag $\tau$ was determined to be $8.0^{+12.6}_{-33.2}$~days ($1\sigma$ errors). 
The error on the negative side is much larger than that on the positive side because both 2--10~keV continuum and narrow Fe-K$\alpha$ data  were sparse after their peaks at MJD~56393--56442.  

We next  applied the same ICCF analysis in the calculation settings to the $B$-band light curve (Figure~\ref{LCs} bottom)   and narrow Fe-K$\alpha$ light curve. 
Figure~\ref{CCF} right-panels show the derived CCF and CCCD of $\tau$.
The result is consistent, within errors, with the ICCF analysis with the 2--10~keV continuum
and $\tau$ was more tightly constrained to be $10.9^{+4.5}_{-9.4}$~days ($1\sigma$ errors), owing to the denser data points of the $B$-band light curve than those of the 2--10~keV continuum, in particular after the flare peak at MJD~56393--56442.
 The $B$-band flux was already confirmed to show  a strong correlation  with the X-ray continuum flux with a small time lag of $\sim 2$~days \citep{2016ApJ...828...78N}.  Hence, we consider that the obtained $\tau$ between the $B$-band and narrow Fe-K$\alpha$ gives a good indication for $\tau$ between the X-ray continuum and narrow Fe-K$\alpha$.           
Table~\ref{tab:lag} summarizes the obtained time lags $\tau$.

%==================================================
\subsection{JAVELIN analyses}
\label{3.3}
%==================================================

%%%%%%%%%%%%%%%%figure4%%%%%%%%%%%%%%%%%%%%%%%%
\begin{figure*}[t]
\epsscale{1.2}
\plotone{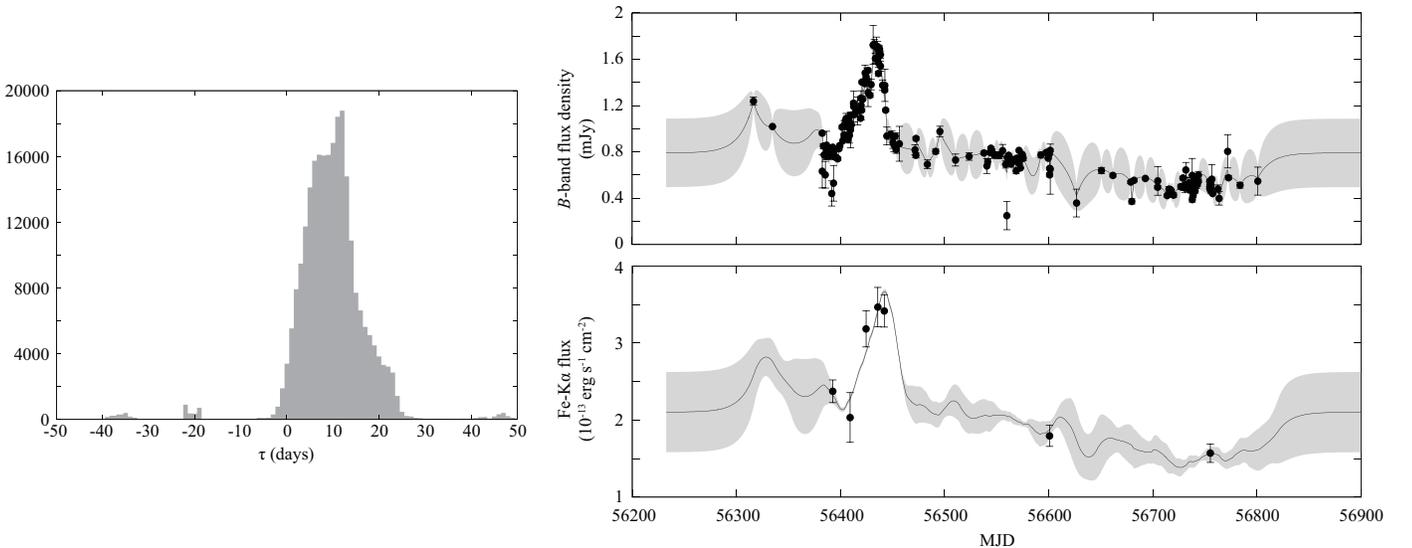}
\caption{Same as Figure~\ref{JAVELIN_suzaku}, but using the $B$-band and narrow Fe-K$\alpha$ light curves as the primary and response light curves, respectively.}
\label{JAVELIN_final}
\end{figure*}
%%%%%%%%%%%%%%%%figure4%%%%%%%%%%%%%%%%%%%%%%%%

In this subsection, we introduce a transfer function which reflects the source geometry of the narrow Fe-K$\alpha$ emission line. 
For this purpose, we conducted the JAVELIN analysis (\citealt{2011ApJ...735...80Z}; \citealt{2013ApJ...765..106Z}).
In the JAVELIN process, the continuum variability is assumed to follow  a damped random walk function, and a simple top-hat shape transfer function is convolved to reproduce the line variability. 
The damped random walk function is described  in terms of the Gauss process with a kernel which includes parameters of  amplitude and the time constant. 
The top-hat transfer function has three parameters, the time lag $\tau$,  width $w$, and  scale $A$. 
The JAVELIN process fits the continuum and line light curves, searching for the best-fit parameters  through the Markov chain Monte Carlo (MCMC) method. 

We performed the JAVELIN analysis using the 2--10~keV continuum and  narrow Fe-K$\alpha$ light curves as the primary and reprocessed components, respectively. 
First, we used the entire period of the 2--10~keV light curve.
Considering the time lag obtained  with the ICCF analyses (Figure~\ref{CCF}) and its uncertainties, we  set a flat prior distribution of $\tau$ from $-50$ to $50$~days. 
If  $w/2$ is larger than  $\tau$, the transfer function violates causality. 
Hence, we assumed the flat prior distribution of $w$ to fall within a range of  $0$ to $30$~days, which is not much larger than the time lag. 
We also assumed a flat prior distribution of $A$ to be from $0$ to $0.10$ to  suppress the possibility of unrealistically large amplitudes of the narrow Fe-K$\alpha$ variation  during periods in which no data points are present. 
Note that $A = 0.10$ corresponds to  an amplitude of $\sim 16$ times larger than the observed one from MJD~56393 to 56436, so  this constraint is not  too tight.
Figure~\ref{JAVELIN_suzaku}~(left) open-bar histogram shows the derived posterior distribution of $\tau$. 
Whereas the posterior distribution showed two  peaks of comparable height around $\tau \sim -30$ and $0$~days, the range of $\tau \gtrsim 30$~days was  ruled out at least.  

Next, in order to focus on the narrow Fe-K$\alpha$ flare at MJD~56393--56442, which was densely covered by \textit{Suzaku}, we limited the period to MJD~56000--56800, i.e.,  a period of roughly $\pm 400$~days from the flare time. 
We again applied the JAVELIN analysis and  showed the obtained posterior distribution of $\tau$  in the shaded histogram in Figure~\ref{JAVELIN_suzaku}~(left). The best-fit light curves are also shown in Figure~\ref{JAVELIN_suzaku}~(right).  
The median  with $1\sigma$ errors  was  $\tau = 4.9^{+10.5}_{-30.3}$~days. 
The uncertainties are relatively large, in particular on the negative side, presumably  
 because of the sparse data after the flare peak (see the previous subsection).     

So far, we  had been using the 2--10~keV flux as the primary X-ray light curve. However, the Fe-K$\alpha$ line  is actually  reprocessed photons of X-rays with  energies above $\sim 7.1$~keV. Hence, ideally, 7--10~keV light-curves should be used, the statistics permitting. 
We calculated the 7--10~keV fluxes from the \textit{Suzaku} and \textit{Swift} spectral  fitting results and performed the JAVELIN analysis between the 7--10~keV continuum and  narrow Fe-K$\alpha$ light curves at the period of MJD~56000--56800.
We obtained a similar posterior distribution of the time lag  of $\tau = 3.2^{+9.2}_{-18.5}$~days (the median \& $1\sigma$ errors). 
Therefore, the 2--10~keV full range can be used  for the energy range of a primary X-ray continuum to correlate with the narrow Fe-K$\alpha$ line as a reasonable substitute, having better statistics and being conventional as the energy range, of the 7--10~keV band continuum. 

Finally, we  conducted the JAVELIN analysis using the $B$-band light curve as the primary continuum, which has better coverage and is reasonable (see discussion in the previous subsection and   \cite{2016ApJ...828...78N}), instead of the 2--10~keV light curve, employing the same calculation settings. 
Figure~\ref{JAVELIN_final} shows  the obtained  posterior distribution of $\tau$ and the best-fit light curve models. 
 The time-lag $\tau$ was  determined to be $\tau = 10.1^{+5.8}_{-5.6}$~days (the median and $1\sigma$ errors). 

Both the JAVELIN analysis results of $\tau$ with the 2--10~keV continuum and $B$-band light-curves  were found to be consistent with those obtained  with the ICCF analyses.
 Table~\ref{tab:lag} summarizes the results. 

%%%%%%%%%%%Table 1%%%%%%%%%%                                                                                                                              
\renewcommand{\arraystretch}{1}
\begin{table}
 \caption{ Time lags in days of the narrow Fe-K$\alpha$  line to the 2--10~keV continuum and $B$-band light curves,  derived with the ICCF and JAVELIN analyses. Medians and $1\sigma$ uncertainties are  tabulated.}
 \label{all_tbl}
 \begin{center}
 \begin{tabular}{ccc}

  \hline\hline
   Primary continuum & \textit{Suzaku}+\textit{Swift}  & $B$ band  \\
          \hline
                            	 
  ICCF ($-300$--300~days) & $8.0^{+12.6}_{-33.2}$ &$10.9^{+4.5}_{-9.4}$ \\
  JAVELIN ($-50$--50~days) &$4.9^{+10.5}_{-30.3}$ &$10.1^{+5.8}_{-5.6}$ \\	
			
  \hline\hline
  \end{tabular}
\end{center}
\label{tab:lag}
\end{table}
%%%%%%%%%%%Table 1%%%%%%%%%%%% 

%==================================================
\subsection{Decoupling of the variable and stationary narrow Fe-K$\alpha$ components}
\label{3.4}
%==================================================

\begin{figure}[t]
\epsscale{1.2}
\plotone{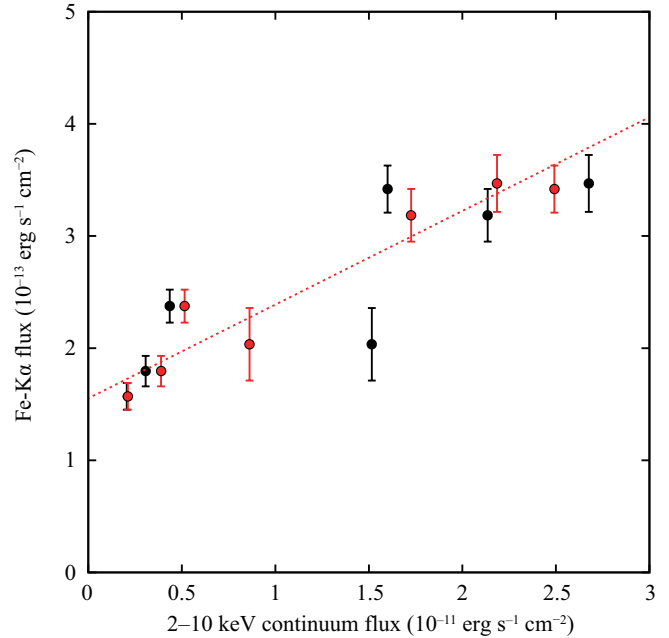}
\caption{ Correlation between the 2--10~keV continuum and narrow Fe-K$\alpha$ fluxes at Epochs~1--7. Black and red show the  data points, in which  the time lags of the Fe-K$\alpha$  to the 2--10~keV continuum flux are assumed to be 0~days and 10.1~days, respectively.  
Red dotted line shows a best-fit linear function obtained by fitting the red data points only considering the errors of narrow Fe-K$\alpha$ fluxes.}
\label{scatter_cont_Fe}
\end{figure}

We distinguish the variable and stationary components in the narrow Fe-K$\alpha$ fluxes. 
Since the JAVELIN algorithm ignores stationary components, we cannot directly  decouple them in the ordinary routine of JAVELIN. 
 Then, we made a  correlation plot between the 2--10~keV continuum flux versus the narrow Fe-K$\alpha$ flux at Epochs~1--7 (Figure~\ref{scatter_cont_Fe}) in the following two procedures. 

First, we  assumed no time lag  between the narrow Fe-K$\alpha$ flux  and the 2--10~keV continuum variation. 
The scatter plot between them is shown in Figure~\ref{scatter_cont_Fe} (black). 
Although it seems to have a marginally positive correlation, some large scatters  remained. 
To test the correlation strength, we calculated the Pearson correlation coefficient. 
As a result, it became 0.84, with which no correlation cannot be ruled out for the number of samples of 7 with the $1\%$ significance level. 
Model fitting of the plot with a linear function $y=ax+b$, where $x$ is the 2--10~keV continuum flux  and $y$ is the Fe-K$\alpha$ flux, yielded an unacceptable result with $\chi^2$/d.o.f.$=24.7/5$. 

  %%%%%%%%%%%%%%%%figure1%%%%%%%%%%%%%%%%%%%%%%%%
%%%%%%%%%%%%%%%%figure1%%%%%%%%%%%%%%%%%%%%%%%%

Next, we made the same plot, but  assuming the time lag of the narrow Fe-K$\alpha$ flux to the 2--10~keV continuum variation of $\tau \sim 10.1$~days, which is the best-fit value derived  with the JAVELIN analysis in \S\ref{3.3}. 
Because the best-fit light-curve model of the 2--10~keV continuum flux obtained by JAVELIN (Figure~\ref{JAVELIN_suzaku} right) includes negative values due to large uncertainties around Epoch~7, it seems unsuited to estimate interpolated fluxes. 
Therefore, we linearly interpolated the 2--10~keV continuum light curve obtained with \textit{Suzaku} and \textit{Swift} (top panel of Figure~\ref{LCs}) and estimated the 2--10~keV continuum fluxes at the times of $10.1$~days before Epochs~1--7, 
whereas the errors on the linearly-interpolated fluxes were not considered.
The narrow Fe-K$\alpha$ fluxes and errors at Epochs~1--7 were simply the observed values at each epoch. 
Figure~\ref{scatter_cont_Fe} (red) shows a scatter plot between the narrow Fe-K$\alpha$ flux and the interpolated 2--10~keV continuum variation.
The Pearson correlation coefficient became 0.96 which indicates a significant positive correlation for the number of samples of 7 with the $1\%$ significance level, showing that the positive correlation strengthened from the case with no time lag. 
For reference, Figure~\ref{scatter_cont_Fe} (red) shows a linear function $y=ax+b$, where $a =  8.4\times10^{-3}$ and  $b = 1.5 \times 10^{-13}$~erg~s$^{-1}$~cm$^{-2}$, which are the best-fit values without considering the errors on the interpolated 2--10~keV fluxes. 
Around Epoch~7, in which the 2--10~keV flux got close to zero, the stationary Fe-K$\alpha$ flux $\sim 1.5 \times 10^{-13}$~erg~s$^{-1}$~cm$^{-2}$ can be confirmed. It remained for a year (from \textit{Suzaku} Epochs~1 to 7) regardless of the 2--10~keV continuum variability during Epochs 1--7. 
At Epoch~3--4,  the variable Fe-K$\alpha$ flux which followed the X-ray continuum variation with the time lag of $\sim 10.1$~days got maximum, and the variable and stationary fluxes became comparable.

%==================================================
\section{Discussion}
\label{four}
%==================================================

%---------------------------------------------------------------------------------------
\subsection{Comparison with previous narrow Fe-K$\alpha$ and multi-wavelength studies}
%---------------------------------------------------------------------------------------

The origin of the narrow Fe-K$\alpha$ emission line has been discussed on many AGNs.  
In the past studies, one or more of the following  three methods were mainly used to investigate the origin: one focusing on the variability of the line flux, one using the line width, and the other directly imaging the spatial distribution of narrow Fe-K$\alpha$ lines. 
In this subsection, we summarize and discuss our results,  comparing ours  with the previous studies in  the methods in the above-listed order.

%------------------------------------------------------------------
\subsubsection{Our results of the narrow Fe-K$\alpha$ line variability}
%------------------------------------------------------------------

The flux variability of the narrow Fe-K$\alpha$ line  provides crucial information to constrain its emitting region,  as several studies have  demonstrated the point. 
The long-term variability on timescales of hundreds to~thousands of days  is common in Seyferts (e.g., \citealt{2016ApJ...821...15F}).  
In the bright type-1 Seyfert NGC~4051, for example,  the narrow Fe-K$\alpha$ flux was reported to  respond possibly to the X-ray continuum flux variation of these timescales  (e.g., \citealt{2003MNRAS.338..323L}). 
The short-term flux variability  in days to~months has been also studied in several Seyferts (e.g., \citealt{2002A&A...388L...5P};  \cite{2013A&A...549A..72P}; \citealt{2015MNRAS.447..160M}; \citealt{2019ApJ...884...26Z}).  
\cite{2013A&A...549A..72P} compared the short-term variation of the narrow Fe-K$\alpha$ flux  to the X-ray continuum variation and argued that its origin  was located at $40$--$1000~R_{\rm g}$ from  the SMBH in  the bright type-1 Seyfert Mrk~509.
\cite{2019ApJ...884...26Z} revisited the archival X-ray  data of the type-1 Seyfert NGC~4151 and argued that the narrow Fe-K$\alpha$ variation had  a time lag of $\sim 3.3$~days behind the X-ray continuum variation. 
Recently, \cite{2022A&A...664A..46A} performed systematic studies combining the flux variability, the Full-Width Half Maximums (FWHM) (\S\ref{4.1.2}) and the direct imaging (\S4.1.3) of the narrow Fe-K$\alpha$ lines and confirmed that the source sizes of the narrow Fe-K$\alpha$  regions are smaller than the dust sublimation radii in most AGNs. 
However, they did not  find the short-term Fe-K$\alpha$ variability in NGC~4151 reported by \cite{2019ApJ...884...26Z}. 
 As such, the understanding of the short-term variability of AGNs  is still  under debate, and further studies are necessary.

%%%%%%%%%%%%%%%%figure4%%%%%%%%%%%%%%%%%%%%%%%%
\begin{figure}[t]
\epsscale{1.1}
\plotone{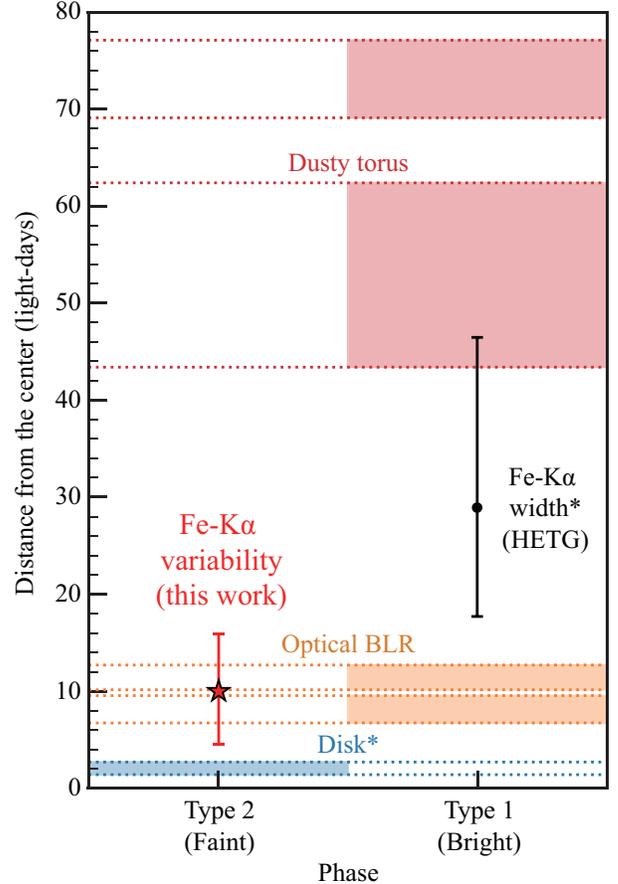}
\caption{Comparison between the distance of the narrow Fe-K$\alpha$ emitting region from the center  in the type-2 (faint) phase in this study  (red star mark) to  previous multi-wavelength studies. 
Blue, orange, and red shaded regions show the $B$-band emitting radius of the accretion disk in the type-2 phase (\citealt{2016ApJ...828...78N}), the distance of the BLR (\citealt{2010ApJ...721..715D}; \citealt{2021ApJ...909...18F}), and dust-reverberation radius in the type-1 phase (JAVELIN and ICCF results of \citealt{2014ApJ...788..159K} from bottom to top), respectively. 
The distance measured  from the width of the narrow Fe-K$\alpha$ line in the type-1 (bright) phase \citep{2010ApJS..187..581S} is also shown  with the filled black circle. 
Note that the distances to the regions with asterisk marks were measured from the SMBH or an X-ray corona,  whereas those without asterisk marks were measured from the optical emitting radius of $\sim 2$~light-days from the X-ray corona.}
\label{Comparison}
\end{figure}
%%%%%%%%%%%%%%%%figure4%%%%%%%%%%%%%%%%%%%%%%%%

In this study, we detected  a significant variability of the narrow Fe-K$\alpha$ line  in NGC~3516 on  a time scale of tens~of~days (Figure~\ref{LCs}). 
This is one of the clearest results  ever obtained in any AGNs about  narrow Fe-K$\alpha$ variations responding to the X-ray continuum.  
 We succeeded in obtaining  its time lag of $\tau \sim 10$~days behind the 2--10~keV and $B$-band variation with two direct methods of the ICCF and JAVELIN. 
The most interesting and novel point of our result is that  the changing-look process occurred and NGC~3516 was in the almost type-2 (faint) phase during our 2013--2014 \textit{Suzaku} monitoring of the source and that we successfully detected the narrow Fe-K$\alpha$ line variability at the faint phase. 

Figure~\ref{Comparison} shows our result  in the type-2 (faint) phase, together with those of previous broad H$\beta$ and dust reverberation studies  while NGC~3516 was in the type-1 (bright) phase. 
Since the radii of the regions emitting broad H$\beta$ lines depend on the luminosities, the previous results were categorized into the following two  observation epochs: one in the past type-1 (bright) phase \citep{2010ApJ...721..715D} and the other  in the type-1 but  relatively low-flux phase  immediately after NGC~3516 started re-brightening from the type-2 (faint) phase \citep{2021ApJ...909...18F}. 
The dust sublimation radii obtained  with near-infrared reverberation mapping with the CCF and JAVELIN methods \citep{2014ApJ...788..159K} are also shown in Figure~\ref{Comparison}.
The disk reverberation time lag  in the type-2 (faint) phase obtained by \cite{2016ApJ...828...78N} is also shown in Figure~\ref{Comparison}. 
Note that although \cite{2016ApJ...828...78N} reported the $B$-band time lag of $2.0^{+0.7}_{-0.6}$~days ($1\sigma$ errors) behind the 2--10~keV band flux, the distance from the center of the narrow Fe-K$\alpha$ source shown in Figure~\ref{Comparison}  does not  take it into account  because it is easier to compare the results of the narrow Fe-K$\alpha$ variation with a time lag behind the $B$ band variation, given that the distances of the BLR and dust reverberation radius were also measured  from the time lags behind the optical variation.    
As a result, the radius of the variable component of the narrow Fe-K$\alpha$ emitting region in the type-2 (faint) phase is found to be consistent with those of the broad H$\beta$ line at the type-1 (bright) phase, and is significantly smaller than the dust reverberation radii.  
We discuss the structure change in NGC~3516  through the changing-look process in \S\ref{4.2}, comparing the results in the type-1 (bright) and type-2 (faint) phases. 
We also discuss the origin of the stationary part of the narrow Fe-K$\alpha$ line in \S\ref{4.1.3}.

%------------------------------------------------------------------
\subsubsection{Previous studies of the narrow Fe-K$\alpha$ line width}
\label{4.1.2}
%------------------------------------------------------------------

The width of an Fe-K$\alpha$ line would reflect the  distance of its source regions from the center of the AGN. 
Studies of the widths of Fe-K$\alpha$ lines from AGNs demand a detector with a high-energy resolution and thus have been predominantly carried out with    the grating spectrometer onboard \textit{Chandra}, the High Energy Transmission Grating (HETG), which has an energy resolution of  $\Delta E/E \sim 30$~eV/6~keV at the energy of the Fe-K$\alpha$ line, as opposed to X-ray CCD cameras, whose energy resolution is limited to $\Delta E/E \sim 120$~eV/6~keV.  
In many studies, the sources of narrow Fe-K$\alpha$ lines  are assumed to have virial motion and their energy spectra have been fitted with simple Gaussian models.

\cite{2004ApJ...604...63Y} analyzed the HETG data of type-1 Seyferts, and reported that the weighted mean of the narrow Fe-K$\alpha$ core widths is $\sim 2400$~km~s$^{-1}$.
\cite{2010ApJS..187..581S} and \cite{2011ApJ...738..147S} systematically studied the line widths of the narrow Fe-K$\alpha$ cores of type-1 and type-2  Seyferts, respectively, by using HETG data, and reported that the mean FWHMs of their samples are  $\sim 2000$~km~s$^{-1}$ for both types. 
They compared  the obtained widths of the Fe-K lines with those of broad H$\beta$ lines and  suggested that their origins are   spatially distributed from  regions  that are a factor of $\sim2$ closer to the SMBH than the BLR to  those parsecs away from  the SMBH, the positions differing from AGN to AGN.
\cite{2015ApJ...802...98M} compared the line widths of the narrow Fe-K$\alpha$ cores with  the dust reverberation results, in addition to the widths of broad H$\beta$ lines,  and concluded that the Fe-K$\alpha$ emitting regions  exist between the broad Balmer line-emitting regions and the points of the dust reverberation radii. 
\cite{2015ApJ...812..113G} concluded that the dust sublimation radii correspond to  the outer envelopes of the narrow Fe-K$\alpha$ line sources in type-1 Seyferts. 

The HETG data of NGC~3516  obtained circa 2000, at which NGC~3516 was  in the type-1 (bright) phase, were also analyzed in the previous studies already (e.g., \citealt{2002ApJ...574L.123T}; \citealt{2010ApJS..187..581S}).  The velocity width of the narrow Fe-K$\alpha$ line was measured to be $3180^{+880}_{-670}$~km~s$^{-1}$ (FWHM), averaging  multiple observation data \citep{2010ApJS..187..581S}. 
In Figure~\ref{Comparison}, we  plot the calculated corresponding radius,  assuming the virial motion around the SMBH with the mass of $4.27 \times 10^7~M_{\odot}$ \citep{2004ApJ...613..682P}. 
In the type-1 (bright) phase of NGC~3516, it cannot be ruled out that the distance from the SMBH to the region emitting the narrow Fe-K$\alpha$ line is same as that emitting broad H$\beta$ lines, because of the large error. 
It is consistent with the dust reverberation radius obtained  with the CCF method, whereas smaller than that obtained with the JAVELIN method reported in \cite{2015ApJ...802...98M} and \cite{2015ApJ...812..113G}.
We note that no observations  have measured the narrow Fe-K$\alpha$ line width of the source  in the type-2 (faint) phase. 

%------------------------------------------------------------------
\subsubsection{Comparison  with previous Fe-K$\alpha$ imaging analyses}
\label{4.1.3}
%------------------------------------------------------------------

 Direct imaging of the Fe-K$\alpha$ line  should give the  tightest constraint  on the narrow Fe-K$\alpha$ source. 
The best angular resolution available in the X-ray band is $\sim 0.5''$, achieved by \textit{Chandra}; accordingly,  the finest scale that can be spatially resolved  is $\sim 10$~pc  for the nearest Seyfert galaxies or more for more distant AGNs.
The first result of extended narrow Fe-K$\alpha$ signals was reported by \cite{2012MNRAS.423L...6M}, and they found $\sim 200$~pc features in the direction of the torus in the nearby Compton-thick AGN (CTAGN), NGC~4945. 
\cite{2014ApJ...791...81A}  imaged the narrow Fe-K$\alpha$ line and hard X-ray continuum in the CTAGN, Circinus Galaxy, and found that they are extended in the direction of the bicone up to several hundreds pc.
 \cite{2019PASJ...71...68K} also reported that some narrow Fe-K$\alpha$ flux is produced at molecular clouds  located $\sim 60$~pc away from the center in Circinus Galaxy, and argued that the fraction of the extended narrow Fe-K$\alpha$ is $\sim 10$\%, while the rest ($\sim 90$\%) is produced in a central region within $\sim 10$~pc. 
\cite{2015ApJ...812..116B} studied the spatial distribution of the narrow Fe-K$\alpha$ line in the nearby CTAGN, NGC~1068, and found that $30\%$ of the line arises from the region broader than $\sim 140$~pc from the center. 
\cite{2021PASJ...73..338N} also reported that the narrow Fe-K$\alpha$ flux is produced at molecular clouds located  several tens of parsecs  away from the center in NGC~1068.
\cite{2017ApJ...842L...4F} discovered the narrow Fe-K$\alpha$ signals extended to a kpc scale in the CTAGN, ESO~428-G014. 
Like these, extended narrow Fe-K$\alpha$ emissions have been found in multiple CTAGNs, while there are no report on type-1 (and changing-look) sources so far. 

In \S\ref{3.4}, we  decoupled the variable and stationary components of the narrow Fe-K$\alpha$ flux of NGC~3516, 
and it is possible that the stationary component corresponds to the extended narrow Fe-K$\alpha$ line beyond a torus scale of a few tens pc.
To check the contribution of the extended emission, we merged and analyzed the \textit{Chandra}/HETG data of 
NGC~3516 (ObsID: 2080, 2431, 2482, 7281, 7282, 8450, 8451, 8452).
Although the HETG data were obtained during the type-1 phase, it can be used to estimate the extended flux. 
First, we made a radial distribution of 6--7~keV counts per pixel from the 0th order data, and compared it with the point spread function (PSF) created by \cite{2021ApJS..257...64K}. 
As a result, the radial distribution did not show significant excess counts from the PSF.    
Next, we made an image from the 0th order data, and extracted a 2--7~keV spectrum from the $2$--$4''$ annular region which corresponds to $\sim 400$--800~pc from the center. 
Here, we set the inner radius to be $2''$, because a region within $2''$ is affected by strong pile-up effects. 
Considering the flux leakage from the point source at the nucleus, we fitted the 2--7~keV spectrum, and estimated the Fe-K$\alpha$ flux to be $4.9^{+3.5}_{-3.3} \times 10^{-15}$~erg~s$^{-1}$~cm$^{-2}$ ($1\sigma$ errors) which is $\sim 30$ times smaller than the stationary component.
Hence, the stationary component is considered to mainly come from a region within $\sim 400$~pc.
Our observations suggest that  the stationary component remained stable  for $\sim 1$~year (Epochs~1--7).  
Then, the lower limit of the size of its emitting region is estimated to be $\sim 0.3$~pc.
We thus conclude that a large part of the stationary component is likely to be produced at  regions located from $\sim 0.3$ to $\sim 400$~pc away from the SMBH, which includes the locations of the dusty torus and the extended molecular region.

%---------------------------------------------------------------------------------------
\subsection{Structure change and the origin of the BLR}
\label{4.2}
%---------------------------------------------------------------------------------------

%---------------------------------------------------------------------------------------
\subsubsection{Structure of the BLR at the type-2 phase}
%---------------------------------------------------------------------------------------

 The disappearance of the broad emission lines  at a (changing-look) transition of the AGN type  from type 1 to 2  means some drastic change in the state of the BLR,  along with a drastic drop in the  mass-accretion rate. 
Two major  scenarios for the BLR state change are  physical disappearance of the BLR materials  and  the deactivation of   the emission-line production  mechanism (see Introduction). 

In this study, we revealed that the radius of the narrow Fe-K$\alpha$ emitting region is $\sim 10$~light-days in the type-2 (faint) phase  (Figure~\ref{Comparison}).  
The radius is consistent with that of the broad H$\beta$ emitting region  in the type-1 (bright) phase. 
This result therefore supports the second  scenario, i.e., the emission-line production was deactivated  in the type-2 phase while the BLR materials remained at the same  locations as  in the type-1 phase. 
The equivalent width (EW)  of the narrow Fe-K$\alpha$ line  is another independent parameter useful to validate the hypothesis because the EW  is a function of  the product of the covering factor and the column density of  the narrow Fe-K$\alpha$ source. 
 Given that EWs are known to depend on X-ray brightness, we should only compare EWs at epochs with similar 2--10~keV fluxes ($F_{2-10}$). 
According to \cite{2010ApJS..187..581S}, NGC~3516 in a type-1 phase showed EW $=83^{+28}_{-26}$~eV at $F_{2-10} = 2.3\times10^{-11}$~erg~s$^{-1}$~cm$^{-2}$ (HETG ObsID:2482). 
 By contrast, the source in a type-2 phase in this study showed EW $=122 \pm 9$~eV  at $F_{2-10} = 2.7\times10^{-11}$~erg~s$^{-1}$~cm$^{-2}$ (Epoch~4 where the source showed the smallest EW and the highest 2--10~keV flux during the \textit{Suzaku} monitoring: Table~\ref{tab:suzaku}). 
This fact likely suggests that neither the covering factor  nor the column density of the materials emitting the narrow Fe-K$\alpha$ line  considerably changed between the type-1 and -2 phases.

Another important question is whether the BLR materials in the type-2 phase, which cannot be detected in optical observations, are in the dust-free gas or dust phases. 
If their degree of ionization/excitation  simply decreases following the rapid UV continuum drops  in the changing-look process, they should remain in the dust-free gas phase. 
However, \cite{2020MNRAS.491.4615K} suggested that in a CLAGN Mrk~590, the inner radius of the dust distribution in the type-2 phase decreased to the radius of the BLR in the type-1 phase in less than $4$~years after  a changing-look process. 
 They interpreted the result in conjunction with  potential new dust formation at the distance of the BLR from the SMBH  on short timescales of years.  
If short-term dust formation commonly  occurs in the turning-off changing-look process, the BLR materials  in the type-2 phase of NGC~3516 might be also in the dust phase. 

The most decisive observational method to distinguish the dust-free gas  and dust phases is  to monitor the variability of the near-infrared flux, which  originates  at the dust sublimation radius (dust reverberation mapping; e.g., \citealt{2014ApJ...788..159K}; \citealt{2019ApJ...886..150M}; \citealt{2020MNRAS.495.2921N}), simultaneously  with monitoring of the X-ray continuum, optical, and narrow Fe-K$\alpha$ emission fluxes.  
Then, if  the narrow Fe-K$\alpha$ time lag is shorter than the dust reverberation time lag, the BLR materials  must be in the dust-free gas, and if longer, in  the dust phase. 

%---------------------------------------------------------------------------------------
\subsubsection{Constraints to the origin of the BLR}
%---------------------------------------------------------------------------------------

Can the origin of the BLR be constrained  with our results of the narrow Fe-K$\alpha$ reverberation mapping?
The following four scenarios are the major ones that have been  proposed  to explain the origin of the BLR, as summarized by \cite{2019OAst...28..200C}: 
(1)  an inflow in which BLR materials come from outer regions, such as dusty tori (e.g., \citealt{2008ApJ...687...78H}; \citealt{2009ApJ...707L..82F}), 
 (2) in-situ formation  through accretion disk fragmentation and star formation (e.g., \citealt{1999A&A...344..433C}; \citealt{2008A&A...477..419C}; \citealt{2011ApJ...739....3W}), 
 (3) the static disk atmosphere puffed up by the dust radiation pressure from  the accretion disk (e.g., \citealt{2018MNRAS.474.1970B}), 
 (4) accretion disk wind which is driven by radiation pressure via lines (e.g., \citealt{1995ApJ...454L.105M}) or dusts (e.g, \citealt{2011A&A...525L...8C}).  
Because of the presence of winds from accretion disks in many AGNs (e.g., \citealt{2013MNRAS.430.1102T}), the last two scenarios are attracting attentions, and hotly debated recently. 

In scenarios (1) and (2), the location of the BLR materials are not directly affected by the luminosity variation in the central engine. Hence, it is difficult to test them by the present results. 
By contrast, in scenarios (3) and (4), the location of the BLR materials is determined  according to the balance between the SMBH gravity and the radiative pressure from  the accretion disk;  then, the location must be affected by the luminosity drop during the changing-look process. 
According to \cite{2018MNRAS.474.1970B} on the third scenario, the peak-height radius of the puffed-up disk, $R_{\rm max}$, which represents the  BLR location, has  a relatively high luminosity dependence  of $R_{\rm max} \propto L^{0.59}$, and accordingly, 
 the BLR radius should decrease by a factor of $\sim 4$ when the luminosity decreases by a factor of $10$.  
Therefore, their prediction is inconsistent with our result that the BLR materials during the type-2 (faint) phase remained at almost the same distance as  in the type-1 (bright) phase. 
However, given that the type-2 phase during which we performed the \textit{Suzaku} monitoring was only $\sim 1$~year after the luminosity decrease  due to the changing-look process started in NGC~3516 at MJD~56000 \citep{2020A&A...638A..13I},  it is possible that  one-year duration was too short  for the geometry of the BLR to change significantly; it  should be on the dynamical timescale. 
To furthermore test scenario (3) and (4), it is  crucial to perform  narrow Fe-K$\alpha$ reverberation mapping on CLAGNs after  both their type-changes and transition of the geometry of the BLR  are completed.

As reported in \cite{2022ApJ...925...84M}, NGC~3516 was observed by \textit{Chandra}/Low Energy Transmission Grating (LETG), \textit{NuSTAR}, and \textit{Swift} almost simultaneously in 2017 during which it was still in the faint phase. 
These are the data obtained $\sim 5$~years after the changing-look process started, and the change of the BLR geometry is considered to have more proceeded. 
Because \textit{Chandra}/LETG does not cover a range above $3$~keV, we focused the \textit{NuSTAR} and \textit{Swift} data. 
We reduced the \textit{NuSTAR} data (ObsID: 603020160$[02, 04, 06, 08, 10,12]$) with similar manners to those in  \cite{2022ApJ...925...84M}, and obtained the 2--10~keV continuum fluxes ($\sim (0.5 - 1.4)\times 10^{-11}$~erg~s$^{-1}$~cm$^{-2}$) and the narrow Fe-K$\alpha$ fluxes ($\sim 2.1 \times 10^{-13}$~erg~s$^{-1}$~cm$^{-2}$) in 2017 Dec by spectral fits with the same model as those in the \textit{Suzaku} spectral fits in \S\ref{3.1}. 
We also analyzed the \textit{Swift} data in MJD~57960.4--58117.5 (2017 Jul--Dec), and obtained the 2--10~keV continuum fluxes from the individual observations by the same manners as in \S\ref{3.1}. 
As a result, the significant variability of the narrow Fe-K$\alpha$ flux was not detected among the \textit{NuSTAR} observations, and   
hence, we cannot constrain the time lag of the narrow Fe-K$\alpha$ flux behind the 2--10~keV flux in 2017 by applying the ICCF or JAVELIN analyses. 
Therefore, it is essential to detect significant variability of the narrow Fe-K$\alpha$ line for further tests of the formation models of the BLR in future monitors.

\subsubsection{Future observations with X-ray microcalorimeters}

The narrow Fe-K$\alpha$ reverberation mapping can be conducted  with  currently operational X-ray CCD and/or grating detectors  in orbit, as we have demonstrated in this study. 
However, observations with X-ray microcalorimeters, which have  unprecedentedly high energy resolution  of $\Delta E/E \lesssim 7$~eV/6~keV, are expected to bring a technical leap in narrow Fe-K$\alpha$ reverberation mapping.   
\cite{2018PASJ...70...13H} performed  ultra-high-quality spectroscopy on the narrow Fe-K$\alpha$ line emission from  the AGN NGC~1275 with the X-ray microcalorimeter onboard \textit{Hitomi}, which was  the first  X-ray microcalorimeter observation of an AGN. 
They detected a very weak (EW $\sim 20$~eV) and narrow (velocity width of 900--1600~km~s$^{-1}$) Fe-K$\alpha$ line, which  could be detected  with neither X-ray CCD cameras  nor grating detectors.  
As such, X-ray microcalorimeters will enable us to detect even modest (or weak) variability of narrow Fe-K$\alpha$ lines. 
The X-Ray Imaging Spectrometer Mission (\textit{XRISM}) with the X-ray microcalorimeter Resolve  is scheduled to be launched  in 2023, and the launch of the Advanced Telescope for High ENergy Astrophysics (\textit{Athena}) with the X-ray microcalorimeter X-IFU is planned  in the 2030s. 
Narrow Fe-K$\alpha$ reverberation mapping with the X-ray microcalorimeters onboard \textit{XRISM} and \textit{Athena} will  reveal the structures of AGNs and their origins in great detail and precision.

%=============================
\section{Conclusions}
%=============================

 With the \textit{Suzaku}, \textit{Swift} and optical simultaneous monitoring observations of the CLAGN NGC~3516 in the almost type-2 (faint) phase, we detected the flux variability of the narrow Fe-K$\alpha$ emission line at $6.4$~keV on  a timescale of tens~of~days, which is correlated  with the 2--10~keV and $B$-band continuum variations.
We performed the narrow Fe-K$\alpha$ reverberation mapping  with the 2--10~keV continuum, $B$ band, and  narrow Fe-K$\alpha$ line light-curves, and succeeded in determining the narrow Fe-K$\alpha$ time lag  to be $10.1^{+5.8}_{-5.6}$~days behind the continuum. 
The narrow Fe-K$\alpha$ emitting radius  in the type-2 (faint) phase is consistent with that of the BLR materials in the type-1 (bright) phase. 
This  implies that even  while the CLAGN was  in the type-2 phase during the changing-look process, the BLR materials remained at the same location as in the type-1 phase and were deactivated from producing optical broad lines  through the rapid drop of the ionizing/exciting UV flux,  which  is presumably caused by the state transition of the accretion flow onto the SMBH. 
 The result might be inconsistent with the hotly-discussed formation models of the BLR which propose that the radiative pressure from the accretion disk should be the main driving force,  given that the radiation  would become too weak to sustain the BLR materials at the same location during the type-2 phase if our interpretation of our result is the case.
This study  demonstrates the efficiency of the narrow Fe-K$\alpha$ reverberation mapping  for revealing the structure of AGNs and its formation mechanism.  The method is applicable in a far more effective way to data obtained with the X-ray microcalorimeters onboard the  scheduled astronomical X-ray satellites \textit{XRISM} and \textit{Athena}.

\acknowledgments

We thank the anonymous referee for his/her valuable suggestions and comments. 
This study is supported by Japan Society for the Promotion of Science (JSPS) KAKENHI with Grant numbers of 19K21884 (HN), 20H00175 (HM), 20H00178 (HM), 20H01941 (HN), 20H01947 (HN), 20K14529 (TK), 22K03683 (HS) and 22K20391(SY).  SY is grateful for support from RIKEN Special Postdoctoral Researcher Program. This work made use of data supplied by the UK Swift Science Data Centre at the University of Leicester.

%\bibliography{template}

%\begin{appendices}
%
%\section{Bayesian Credible Region}
%\label{sec:CredRegion}
%
%In the Bayesian context, given the posterior probability density,
%$p(\theta|{\bf D})$, of a parameter, $\theta$, given the dataset,
%${\bf D}$, the highest posterior density region (or credible
%region), $R$, is defined by
%\begin{equation}
%C = \int\sb{R} {\rm d}\theta\;p(\theta|{\bf D})
%\label{eq:CredRegion}
%\end{equation}
%where C is the probability contained in the credible region.  The
%region $R$ is selected such that the posterior probability of any
%point inside $R$ is larger than that of any point outside.  We like
%Eq.\ (\ref{eq:CredRegion}).
%
%\end{appendices}

\end{document}